\documentclass[%
 reprint,
superscriptaddress,
 amsmath,amssymb,
 aps,
prx,
floatfix,
]{revtex4-1}

\usepackage{ifpdf}
\usepackage{amsmath} 
\usepackage{amssymb}
\usepackage{amsfonts}
\usepackage{braket}
\usepackage{enumitem}
\usepackage{physics}
\usepackage{float}
\usepackage{color}
\usepackage{booktabs}
\usepackage[colorlinks=true,linkcolor=blue,citecolor=blue,breaklinks]{hyperref}
\usepackage{graphicx}
\usepackage{dcolumn}
\usepackage{bm}
\usepackage[normalem]{ulem}
\usepackage{mathtools}
\usepackage{verbatim}
\usepackage{xcolor}

\newcommand{\hc}{\text{H.c.}}

\usepackage{tikz}
\usetikzlibrary{math}
\definecolor{Zcolour}{rgb}{0.992, 0.588, 0.22}
\definecolor{purple}{rgb}{0.5,0,0.5}
\definecolor{brown}{rgb}{0.6,0.2,0}
\definecolor{dkgreen}{rgb}{0,0.5,0}

\begin{document}


\title{Unified structure for exact towers of scar states in the Affleck-Kennedy-Lieb-Tasaki and other models}
\author{Daniel K.~Mark}
\affiliation{Department of Physics, California Institute of Technology, Pasadena, California 91125, USA}
\author{Cheng-Ju Lin}
\affiliation{Perimeter Institute for Theoretical Physics, Waterloo, Ontario N2L 2Y5, Canada}
\author{Olexei I.~Motrunich}
\affiliation{Department of Physics, California Institute of Technology, Pasadena, CA 91125, USA}

\date{\today}

\begin{abstract}
Quantum many-body scar states are many-body states with finite energy density in nonintegrable models that do not obey the eigenstate thermalization hypothesis. Recent works have revealed ``towers" of scar states that are exactly known and are equally spaced in energy, specifically in the AKLT model, the spin-1 XY model, and a spin-1/2 model that conserves the number of domain walls. We provide a common framework to understand and prove known exact towers of scars in these systems, by evaluating the commutator of the Hamiltonian and a ladder operator. In particular we provide a simple proof of the scar towers in the integer-spin 1D AKLT models by studying two-site spin projectors. Through this picture we deduce a family of Hamiltonians that share the scar tower with the AKLT model, and also find common parent Hamiltonians for the AKLT and XY model scars.
We also introduce new towers of exact states, organized in a ``pyramid" structure, in the spin-1/2 model through the successive application of a nonlocal ladder operator.
\end{abstract}

\maketitle

\section{Introduction}
The eigenstate thermalization hypothesis (ETH) is a paradigm for thermalization in closed quantum many-body systems~\cite{Deutsch1991, Srednicki1994}. It is a framework to describe how nonintegrable quantum many-body systems equilibrate to thermal ensembles consistent with statistical mechanics. A \emph{strong} ETH appears to hold in many systems, where \emph{all} eigenstates at finite energy density obey the ETH. Many-body localized systems \cite{Basko2006, Oganesyan2007:MBLhighT, Bardarson2012:UnboundedEEgrowthMBL, Serbyn2013:SlowEEgrowth, serbyn_local_2013, Huse2014:PhenomMBL, Bauer2013:AreaLawMBL, nandkishore_many-body_2015, Abanin2019:RMPonMBL}
strongly violate the ETH. A recent experiment~\cite{Bernien2017} on cold Rydberg atoms observed unusual quench dynamics in a nonintegrable many-body system. This was attributed to a set of special eigenstates and set off a flurry of work~\cite{turner_quantum_2018, Turner2018, khemani_signatures_2018, ho_periodic_2019, Lin2019, Choi2019, surace_lattice_2019, Iadecola2019, bull_systematic_2019, James2019:Confinement, Pai2019:ScarsFracton, Sala2019:Fragmentation, 
Shiraishi2019:PXPembeddedH,
michailidis_slow_2019, moudgalya_quantum_2019, lin_slow_2019, khemani_local_2019, hudomal_quantum_2019, pancotti_quantum_2019, alhambra_revivals_2019} on so-called ``quantum many-body scar states," which are eigenstates of nonintegrable Hamilitonians that do not obey the ETH. This is in analogy to single-particle scar states~\cite{heller_bound-state_1984} in an otherwise chaotic spectrum.
While these many-body systems do not obey the \emph{strong} ETH, the \emph{weak} ETH holds; that is, the ETH holds for \emph{almost} every eigenstate. The ``PXP model," which describes the original Rydberg atom experiment, was extensively studied~\cite{Fendley2004:HardBosonModel, Bernien2017, turner_quantum_2018, Turner2018, khemani_signatures_2018, ho_periodic_2019, Lin2019, Choi2019, surace_lattice_2019, Iadecola2019, Shiraishi2019:PXPembeddedH, michailidis_slow_2019, moudgalya_quantum_2019, lin_slow_2019,bull_quantum_2020}. While there are several approximation schemes~\cite{turner_quantum_2018, Turner2018, ho_periodic_2019, Lin2019, surace_lattice_2019, Iadecola2019,bull_quantum_2020} to understand the scar states in the PXP model, the scars are not exactly known (except for some eigenstates in the middle of the spectrum~\cite{Lin2019, Shiraishi2019:PXPembeddedH}). 

In contrast, there are nonintegrable systems with exactly known scar states~\cite{yang__1989,shiraishi_systematic_2017,Mori2017:ThermalizationWithoutETH,vafek_entanglement_2017,Moudgalya2018,moudgalya_entanglement_2018,schecter_weak_2019,iadecola_quantum_2019,chattopadhyay_quantum_2019,shibata_onsagers_2019}. In particular, several models are known to host ``towers" of exact scar states, where the scar states are known analytically and are equally spaced in energy. They are also frequently obtained through successive application of an operator $Q^\dagger$ on some initial state. 

Reference~\cite{Moudgalya2018} introduced a tower of exact states in the Affleck-Kennedy-Lieb-Tasaki (AKLT) model~\cite{affleck_rigorous_1987}, which were subsequently shown in Ref.~\cite{moudgalya_entanglement_2018} to have sub-volume entanglement entropy. The spin-1 AKLT model is of theoretical importance for having an exactly known gapped ground state in the Haldane phase. The ground state exhibits symmetry protected topological (SPT)~\cite{pollmann_symmetry_2012} order and can be expressed as a bond dimension two matrix product state (MPS)~\cite{klumper_matrix-product-groundstates_1993}.

In Ref.~\cite{schecter_weak_2019}, Schecter and Iadecola introduced a tower of exact states in a family of nonintegrable spin-1 XY-type models. In Ref.~\cite{iadecola_quantum_2019}, Iadecola and Schecter also introduced a tower of exact states in a particular nonintegrable spin-1/2 model that conserves number of domain walls. In both cases, the towers of states have sub-volume law entanglement, and Iadecola and Schecter presented an initial quench state to achieve perfect revivals, in which the state periodically returns to itself during time evolution.

In the above three models, the exact towers of states can be written in the form $\ket{\Psi_n} = (Q^\dagger)^n \ket{\Psi_0}$, where $Q^\dagger$ is an operator and $\ket{\Psi_0}$ is a known eigenstate. In this paper we provide alternate proofs for these towers by showing that for every scar state $\ket{\Psi_n}$, $\left[H, Q^\dagger \right] \ket{\Psi_n} = \omega Q^\dagger\ket{\Psi_n}$, which immediately gives us that the $\ket{\Psi_n}$ are eigenstates of $H$, equally spaced in energy by $\omega$. This highlights a common motif and we state a simple theorem.

\textbf{Theorem:} Suppose we have a Hamiltonian $H$; a linear subspace $W$; a state $\ket{\Psi_0} \in W$, which is an eigenstate of $H$ with energy $E_0$; and an operator $Q^\dagger$ such that $Q^\dagger W \subset W$ and
\begin{equation}
\left(\left[H, Q^\dagger \right] - \omega Q^\dagger \right) W = 0 ~.
\label{eq:asimpletheorem}
\end{equation}
Then $(Q^\dagger)^n \ket{\Psi_0}$, as long as it is a nonzero vector, is an eigenstate of $H$ with eigenvalue $E_0 + n \omega$. 

We note that the $\eta$-pairing states in the Hubbard model on bipartite lattices~\cite{yang__1989} provide a special example of this structure, where the subspace $W$ is the entire Hilbert space and $Q^\dagger$ essentially corresponds to a symmetry of the Hamiltonian~\cite{Yang_SO4_1990, Zhang_Pseudospin_symmetry_1990}. In the scar case, $W$ is not the entire space and there is no symmetry associated with $Q^\dagger$.

In the proofs of specific towers, $H$, $\ket{\Psi_0}$, and $Q^\dagger$ are fixed, while for $W$ we can choose the space spanned by $\ket{\Psi_0}, Q^\dagger \ket{\Psi_0}, \dots, (Q^\dagger)^n \ket{\Psi_0}, \dots$, and verify that $[H, Q^\dagger] - \omega Q^\dagger$ annihilates these states.  
Often the null space of $[H, Q^\dagger] - \omega Q^\dagger$ is much larger, and it is easy to find a large subspace $W$ containing $\ket{\Psi_0}$ and preserved by the action of $Q^\dagger$ (the subspace $W$ can be thought of as a property of the states in the tower, which is easy to see in the $\ket{\Psi_0}$ and whose preservation under $Q^\dagger$ is easy to verify).

In Sec.~\ref{sec:spin1AKLT}, we briefly review the scar tower in the AKLT model and provide a short proof for it, relying on spin-projectors on every two neighboring sites. This picture allows us to immediately provide a family of Hamiltonians that share these scar states, and we discuss its possible relation to the Shiraishi-Mori embedded Hamiltonian structure introduced in Ref.~\cite{shiraishi_systematic_2017}.

In Sec.~\ref{sec:spin2AKLT}, we adapt the argument to the 1D spin-2 AKLT model. The two-site projector framework allows us to immediately generalize our proof to higher-spin AKLT models in 1D. We also discuss how the argument fails in higher dimension.

In Sec.~\ref{sec:spin1XY}, we apply the commutator framework discussed above to prove the scar tower in the spin-1 XY model. This scar tower is produced by the same scar operator $Q^\dagger$ as in the spin-1 AKLT model. In Section~\ref{sec:parentham}, we discuss a ``parent Hamiltonian" which is embedded in both the spin-1 XY and AKLT models. This model exhibits $SU(2)$ symmetry generated by $Q^\dagger$ and its conjugate $Q$. We also introduce a family of $SU(2)$-invariant models that share both sets of scar towers.

In Sec.~\ref{sec:spin1/2model}, we study the domain-wall-conserving spin-1/2 model discussed by Iadecola and Schecter. We introduce new exact towers of states, arranged as a ``pyramid" of states, which are related by a nonlocal ladder operator. Lastly, in Sec.~\ref{sec:perfectrevivals}, we discuss ``perfect revivals" of some initial states in these scarred models.   

The common motif of commutators $\left[H, Q^\dagger \right]$ provides a quick way to prove exact towers of states and sheds light on the structure of these scar towers.

\section{The Spin-1 AKLT model}
\label{sec:spin1AKLT}
\subsection{Hamiltonian}
The spin-1 AKLT model~\cite{affleck_rigorous_1987} is a spin-1 1D chain with the following Hamiltonian:
\begin{equation}
    H = \sum_{j=1}^L \left(\frac{1}{3} + \frac{1}{2} \vec{S}_j \cdot \vec{S}_{j+1} + \frac{1}{6} (\vec{S}_j \cdot \vec{S}_{j+1})^2 \right) ~.
\end{equation}
We assume periodic boundary conditions (PBC), i.e., $L+1 \equiv 1$. In this paper we will also assume $L$ even. (We can generalize results also to open boundary conditions (OBC), where the upper limit of the sum is $j=L-1$ instead). The AKLT model is equivalently written in terms of spin-2 projectors:
\begin{equation}
    H = \sum_{j=1}^L P^{(2,1)}_{j,j+1} = \sum_{j=1}^L \left(\sum_{M = -2}^2 \ketbra{T_{2,M}}{T_{2,M}} \right)_{j,j+1} ~,
\end{equation}
where, following the notation of Ref.~\cite{Moudgalya2018}, $P^{(2,1)}_{j,j+1}$ is the projector of two spin-1's on sites $j$, $j+1$ onto total spin-2.
$\ket{T_{J,M}}$ denotes a two-site state of total spin $J$ and $z$ component $M$. For reference we list these states in Appendix~\ref{sec:twositestates}.

The AKLT model is known to be nonintegrable. Despite this, the ground state, several low-lying and highly excited states, and a tower of low-entanglement ``scar states" are known exactly and reviewed in Ref.~\cite{Moudgalya2018}.

\subsection{Ground state}
The ground state of the AKLT model is known exactly~\cite{affleck_rigorous_1987}. It can be compactly expressed by the MPS~\cite{moudgalya_entanglement_2018}:
\begin{equation}
    \ket{G} = \sum_{\{\sigma_1 \cdots \sigma_L\}} \Tr(A^{[\sigma_1]} \cdots A^{[\sigma_L]}) \ket{\sigma_1 \cdots \sigma_L} ~,
\end{equation}
where
\begin{align}
    A^{[-1]} &= \sqrt{\frac{2}{3}} \begin{pmatrix} 0 & 1\\ 0 & 0 \end{pmatrix} ~, 
    \quad A^{[0]} = \frac{1}{\sqrt{3}} \begin{pmatrix} -1 & 0\\ 0 & 1 \end{pmatrix} ~, \nonumber \\ 
    A^{[1]} &= \sqrt{\frac{2}{3}} \begin{pmatrix} 0 & 0\\ -1 & 0 \end{pmatrix} ~.
\label{eq:MPS}
\end{align}
Crucially, $\ket{G}$ is the unique state that does not have any spin-2 component on any bond $(j, j+1)$ and hence has energy $0$.

\subsection{Exact scar states}
\label{sec:spin1AKLTexactscarstates}
The low-entanglement scar states $\ket{\mathcal{S}_{2n}}$ are constructed in Ref.~\cite{Moudgalya2018} atop the ground state. They are defined, up to normalization constants:
\begin{gather}
    \ket{\mathcal{S}_{2n}} = (Q^\dagger)^n \ket{G} ~, \nonumber\\ 
    Q^\dagger = \sum_{j=1}^L (-1)^j (S^+_j)^2 ~, ~~~ n=0,...,L/2 ~.
\end{gather} 
These states span the ground state ($n=0$) to the ferromagnetic state $\ket{1,1...,1}$ ($n=L/2$).
(If $L/2$ is odd, then $(Q^\dagger)^{L/2} \ket{G} = 0$, so the tower does not reach the ferromagnetic state, but this is a small technicality.) These states have energy $E = 2n$ and were shown in Ref.~\cite{moudgalya_entanglement_2018} to have subvolume entanglement entropy $S \propto \ln \ell$ (for subsystem size $\ell$) and were argued to violate the strong ETH. It is also worth noting that these states are not unique: $\ket{\mathcal{S}_{2n}}$ has total spin $s=2n$, and there are equivalent spin-rotated scar states owing to the SU(2) symmetry in the AKLT model.

The original proof of the scar states $\ket{\mathcal{S}_{2n}}$ used a dimer picture and relied on cancellation of scattering dimer configurations under the action of $H$. In what follows we will present a simple proof in the spin basis.

\subsection{Short proof of scar states}
We prove the scar states as follows. For every $n$, to prove that $\ket{\mathcal{S}_{2n}}$ is an eigenstate of $H$ with energy $2n$, it suffices to show that
\begin{equation}
    [H, Q^\dagger] \ket{\mathcal{S}_{2n}} = 2 Q^\dagger \ket{\mathcal{S}_{2n}} ~,
\label{eq:comm1}
\end{equation}
for every $n=0,...,L/2$. That is, the commutator $[H, Q^\dagger]$ is equivalent to the operator $2Q^\dagger$ on the space spanned by the scar states. 

We prove this by showing that
\begin{equation}
    [H, Q^\dagger] = 2 Q^\dagger + A ~, ~~~ A \ket{\mathcal{S}_{2n}} = 0 ~,
\label{eq:comm2}
\end{equation}
for all $n$. To do so we write the commutator as
\begin{equation}
\label{eq:commutator}
    [H, Q^\dagger] = \sum_{j=1}^L (-1)^j \left[P^{(2,1)}_{j,j+1}, (S_j^+)^2 - (S_{j+1}^+)^2 \right] ~.
\end{equation}
After some computation, the operator $(S_j^+)^2 - (S_{j+1}^+)^2$ can be written as
\begin{align}
\label{eq:qj}
    & (S_j^+)^2 - (S_{j+1}^+)^2 =  2\Big(-\ketbra{T_{2,1}}{T_{1,-1}}\\
    & -\sqrt{2}\ketbra{T_{2,2}}{T_{1,0}} + \ketbra{T_{1,1}}{T_{2,-1}} + \sqrt{2}\ketbra{T_{1,0}}{T_{2,-2}} \Big)_{j,j+1} ~. \nonumber
\end{align}
Here we have expressed the operator in terms of the two-site states $\ket{T_{J, M}}$ listed in Eq.~(\ref{eq:twositestates}).
The action of this operator is visualized in Fig.~\ref{fig:qj}. This expression is useful because in the commutator Eq.~(\ref{eq:commutator}), the projector simply projects out different terms in Eq.~(\ref{eq:qj}):
\begin{figure}
    \centering
\includegraphics[width=0.5\textwidth]{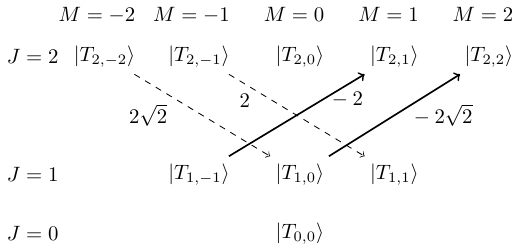}
    \caption{Action of operator $(S_j^+)^2 -(S_{j+1}^+)^2$, Eq.~(\ref{eq:qj}), on the states $\ket{T_{J, M}}_{j,j+1}$ on sites $j$, $j+1$ with total spin $J$ and magnetization $M$.
    }
    \label{fig:qj}
\end{figure}
\begin{align}
    &\left[P^{(2,1)}_{j,j+1}, (S_j^+)^2 -(S_{j+1}^+)^2 \right] = -2 \Big(\ketbra{T_{2,1}}{T_{1,-1}} \\ 
    & +  \sqrt{2}\ketbra{T_{2,2}}{T_{1,0}} + \ketbra{T_{1,1}}{T_{2,-1}}
    +\sqrt{2}\ketbra{T_{1,0}}{T_{2,-2}} \Big)_{j,j+1} ~. \nonumber
\end{align}
We rewrite this as
\begin{gather}
    \left[P^{(2,1)}_{j,j+1}, (S_j^+)^2 - (S_{j+1}^+)^2 \right] = (S_j^+)^2 - (S_{j+1}^+)^2 \nonumber\\
    -4\left(\ketbra{T_{1,1}}{T_{2,-1}} + \sqrt{2}\ketbra{T_{1,0}}{T_{2,-2}} \right)_{j,j+1} ~.
\end{gather}
The terms in the second line annihilate the scar states $\ket{\mathcal{S}_{2n}}$, noting that:
\begin{align}
    &\braket{T_{2,-1}}{\mathcal{S}_{2n}}_{j,j+1} = \frac{1}{\sqrt{2}} \left(\bra{0,-1} + \bra{-1,0} \right)_{j,j+1} \ket{\mathcal{S}_{2n}} = 0 ~, \nonumber\\
    &\braket{T_{2,-2}}{\mathcal{S}_{2n}}_{j,j+1} = \bra{-1,-1}_{j,j+1} \ket{\mathcal{S}_{2n}} = 0 ~.
    \label{eq:S2-2}
\end{align}
This gives the desired relation Eq.~(\ref{eq:comm2}) and completes the proof of the exact scar tower.

We can deduce Eq.~(\ref{eq:S2-2}) either from a direct wavefunction picture in terms of the spin-1's or from the two-site spin-2 projector picture. At the spin-1 level, note that the MPS of the ground state Eq.~(\ref{eq:MPS}) exhibits ``string order": `+1'/`-1' can be followed by any number of `0's, but must be followed by a `-1'/`+1'. In other words, any nonzero spin configuration in $\ket{G}$ must have the pattern ``$\pm1, \mp1, \pm1, ..., \mp1$", with any number of `0's in between. Furthermore, $A^{[-1]} A^{[0]} = -A^{[0]} A^{[-1]}$ and the sequences ``0,-1" and ``-1,0" occur with opposite phases, i.e., the ground state is annihilated by $\ket{...}\left(\bra{0,-1} + \bra{-1,0} \right)_{j,j+1}$.

Since the operator $Q^\dagger$ sends `-1's to `+1's, the scar states can only have patterns of `+1's and `-1's such that `-1's occur in ones, and `+1's occur in bunches of odd length, again with any number of `0's in between. We then see that the sequence ``-1,-1" can never occur, and since only `+1's are produced by $Q^\dagger$, the sequences ``0,-1" and ``-1,0" still occur with opposite phases. This gives the desired relations in Eq.~(\ref{eq:S2-2}).

We can also derive Eq.~(\ref{eq:S2-2}) from the two-site spin-2 projector picture. The ground state $\ket{G}$ has energy 0 and therefore every two-site spin state has either $J=0$ or $J=1$. When $Q^\dagger$ is applied to $\ket{G}$, $\ket{T_{1,-1}}, \ket{T_{1,0}}$ can transition to $\ket{T_{2,1}}$ and $\ket{T_{2,2}}$ respectively. Successively applying $Q^\dagger$, we see that the states $\ket{T_{2,1}}$ and $\ket{T_{2,2}}$ are the only $J=2$ states present across any bond in $\ket{\mathcal{S}_{2n}}$. This gives the desired relations Eq.~(\ref{eq:S2-2}) and additionally $\braket{T_{2,0}}{\mathcal{S}_{2n}}_{j,j+1} = 0$. This last statement can also be verified in the spin-1 picture.

Finally, we note that in this argument, we did not need to know the coefficients $\pm2$ and $\pm2\sqrt{2}$ in the expression for $(S_j^+)^2 - (S_{j+1}^+)^2$ in Eq.~(\ref{eq:qj}). We only needed to know which transitions are present. This can be obtained from the following selection rules:
\begin{enumerate}
    \item The magnetization $M$ increases by 2.
    \item The parity of $J$ changes, because $(S_j^+)^2 - (S_{j+1}^+)^2$ is odd under exchange $j \leftrightarrow j+1$.
\end{enumerate}
One can easily verify that the only allowed transitions are those in Eq.~(\ref{eq:qj}) and Fig.~\ref{fig:qj}, and this is sufficient to prove the exact tower of scar states. This paves the way for generalization to higher-spin 1D AKLT models. 

\subsection{Family of Hamiltonians sharing the tower of states}
\label{subsec:familyS1}
The above picture easily suggests generalizations of the spin-1 AKLT model that share the same tower of scar states. The family of models
\begin{gather}
    H_{\text{new}}^{(1)}
    = \sum_{j=1}^L P^{(2,1)}_{j,j+1} + \Big( a\ketbra{T_{2,-2}}{T_{2,-2}} + b\ketbra{T_{2,-1}}{T_{2,-2}}  
    \nonumber\\ 
    + c\ketbra{T_{2,0}}{T_{2,-2}} + d\ketbra{T_{2,-1}}{T_{2,-1}} + 
    e\ketbra{T_{2,0}}{T_{2,-1}}
    \nonumber\\ 
    + f\ketbra{T_{2,0}}{T_{2,0}} + \hc\Big)_{j,j+1}
    \label{eq:scarfamily}
\end{gather}
is a six-parameter family of Hamiltonians that share the scar states $\ket{\mathcal{S}_{2n}}$. This can be verified by noting that the additional terms annihilate $\ket{\mathcal{S}_{2n}}$, and therefore the scar states are eigenstates of $H'$. 

These terms can be used to systematically break the symmetries in the AKLT model: the diagonal terms $\ketbra{\mathcal{S}_{2,m}}{\mathcal{S}_{2,m}}$ break the SU(2) symmetry, while the off-diagonal terms $\ketbra{\mathcal{S}_{2,m}}{\mathcal{S}_{2,l}},~l\neq m$ break the U(1) symmetry that corresponds to conservation of the total magnetization $\sum_j S_j^z$.
One can also introduce site-dependent coefficients $(a_j, b_j, ..., f_j)$ to break the lattice translation symmetry. 
This can be useful, for example, to generate Hamiltonians that definitively violate the strong ETH. The original 1D AKLT model is
only argued numerically to do so, because it is not known analytically whether the scar states have finite energy density in their respective symmetry sectors. Breaking many symmetry sectors can provide known ground and highest excited states and produce systems that definitively violate the strong ETH.

\subsection{Possible Relation to a Generalization of the Shiraishi-Mori Form}

The above family of Hamiltonians in fact has some similarity to the form discussed by Shiraishi and Mori in Ref.~\cite{shiraishi_systematic_2017}. They proposed a general mechanism to embed ETH-violating eigenstates in an otherwise nonintegrable Hamiltonian.

This is done through local projection operators $P_j$ associated with each site (the operators themselves can act on a few nearby spins), and a subspace $\mathcal{T}$ of the Hilbert space such that $P_j \mathcal{T} = 0$. Then for any Hamiltonian $H'$ such that
\begin{equation}
    \left[H', P_j \right] = 0 ~,
    \label{eq:shirasihimoricomm}
\end{equation}
the family of Hamiltonians
\begin{equation}
    H = \sum_j P_j h_j P_j + H'
\end{equation}
has $\dim \mathcal{T}$ special eigenstates---namely, eigenstates of $H'$ restricted to $\mathcal{T}$---which can be tuned to be in the middle of the energy spectrum. Here $h_j$ can be an arbitrary local Hamiltonian, making the full $H$ in general nonintegrable.

It is tempting to write the family Eq.~(\ref{eq:scarfamily}) in a similar form with
\begin{align}
P_j &= \left(\ketbra{T_{2,-2}}{T_{2,-2}} + \ketbra{T_{2,-1}}{T_{2,-1}} + \ketbra{T_{2,0}}{T_{2,0}} \right)_{j,j+1} ~, \nonumber\\
H' &= \sum_j \left(\ketbra{T_{2,1}}{T_{2,1}} + \ketbra{T_{2,2}}{T_{2,2}} \right)_{j,j+1} ~.
\end{align}
However, in this case, the condition Eq.~(\ref{eq:shirasihimoricomm}) is not satisfied, through terms
\begin{gather}
    \left[ \ketbra{T_{2,m}}{T_{2,m}}_{j-1,j}, \ketbra{T_{2,n}}{T_{2,n}}_{j,j+1} \right] \neq 0~,\\
    \text{for pairs}~~(m,n) = (1,-1), (1,0), (2,0) ~. \nonumber
\end{gather}
Further, in the sums over terms in $H'$, no cancellation occurs to restore the commutation relation.
Finally, the common null space of the above $P_j$'s contains more states than just the scar tower $\ket{\mathcal{S}_{2n}}$.
Thus, the above writing by itself does not reveal the scar states, and we need to resort to the previous proofs for them.
At present, we do not know if it is possible to find a set of local projectors whose common null space would give precisely the tower $\ket{\mathcal{S}_{2n}}$ and which could be used to cast the AKLT chain in the Shiraishi-Mori form.
While this is not a proof, it suggests the possibility of larger families of models with scar states that go beyond the Shiraishi-Mori form.

\section{The Spin-2 AKLT model and generalizations}
\label{sec:spin2AKLT}
\subsection{Hamiltonian}
The spin-2, 1D AKLT model is given by:
\begin{equation}
    H^{(2)} = \sum_{j=1}^L \left( P^{(3,2)}_{j,j+1} + P^{(4,2)}_{j,j+1} \right) ~,
\end{equation}
where $P^{(J,2)}_{j,j+1}$ are projectors onto states of total spin $J$, formed by spin-2's on sites $j$ and $j+1$.

The ground state, denoted $\ket{2G}$, can be expressed through Schwinger bosons~\cite{Moudgalya2018}, but we will only need the fact that it has energy 0 and therefore has no two-site states with spin 3 or 4.

\subsection{Exact scar states and proof}
\label{subsec:proofspin2AKLT}
\begin{figure*}[t]
    \centering
\includegraphics[width=\textwidth]{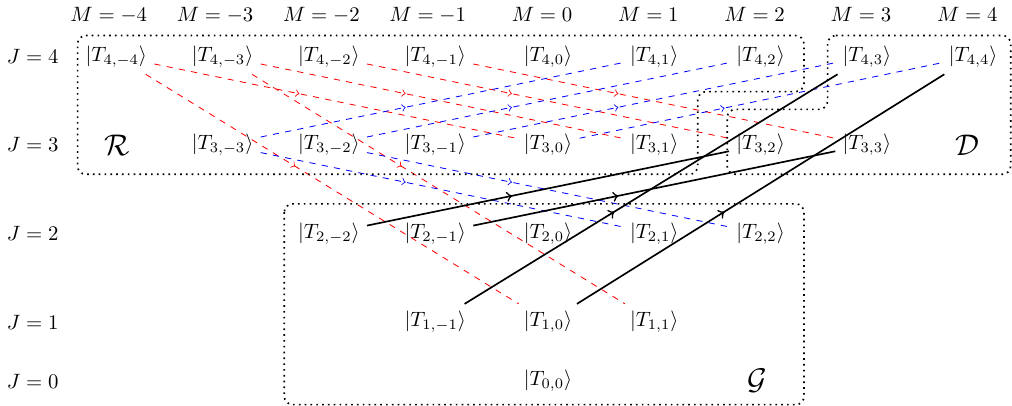}
    \caption{Action of operator $(S_j^+)^4 - (S_{j+1}^+)^4$, Eq.~(\ref{eq:2qj}), on the states $\ket{T_{J,M}}_{j,j+1}$ on sites $j$, $j+1$ with total spin $J$ and $z$-axis magnetization $M$. To prove the desired commutation relation in Eq.~(\ref{eq:Spin2comm}), the states are grouped (surrounded by dotted lines) into $\mathcal{G}$, $\mathcal{D}$, and $\mathcal{R}$ based on whether they are present in the ground state, present in $\ket{2\mathcal{S}_{2n}}$, and never present in $\ket{2\mathcal{S}_{2n}}$ respectively. The corresponding schematic matrix form of this operator is given in Eq.~(\ref{eq:qmat}).}
    \label{fig:spin2q}
\end{figure*}
Ref.~\cite{Moudgalya2018} also derived an exact tower of states for the spin-2 AKLT model. They are, up to normalization constants,
\begin{gather}
    \ket{2\mathcal{S}_{2n}} = (Q^\dagger)^n \ket{2G},\nonumber\\
    Q^\dagger = \sum_{j=1}^L (-1)^j (S^+_j)^4 ~, ~~~ n=0,...,L/2 ~, \label{eq:2qj}
\end{gather}
with energies $E = 2n$ and sub-volume-law entanglement entropy. (The notation is from Ref.~\cite{Moudgalya2018}, where $\ket{SG}$ denotes the ground state of the spin-$S$ AKLT chain and $\ket{S\mathcal{S}_{2n}}$ denotes the corresponding tower state with energy $2n$.)

To prove these scar states we repeat our derivation for the spin-1 scars. We want to show that
\begin{equation}
    [H^{(2)}, Q^\dagger] \ket{2\mathcal{S}_{2n}} = 2 Q^\dagger \ket{2\mathcal{S}_{2n}}.
\label{eq:Spin2comm}
\end{equation}
As above we write:
\begin{align}
[H^{(2)}, Q^\dagger] &= \sum_{j=1}^L (-1)^j \left[P^{(3,2)}_{j,j+1} + P^{(4,2)}_{j,j+1},  (S_j^+)^4 - (S_{j+1}^+)^4 \right] \nonumber \\
&= \sum_{j=1}^L (-1)^j \left[h_{j,j+1}, q_{j,j+1} \right] ~,
\end{align}
where $h_{j,j+1} = P^{(3,2)}_{j,j+1} + P^{(4,2)}_{j,j+1}$ and $q_{j,j+1} = (S_j^+)^4 - (S_{j+1}^+)^4$. To deduce the action of $q_{j,j+1}$, we use the following selection rules:
\begin{enumerate}
    \item The magnetization $M$ increases by 4.
    \item The parity of $J$ changes, because $q_{j,j+1}$ is odd under exchange $j \leftrightarrow j+1$.
\end{enumerate}
We do not calculate the coefficients, but display all the allowed transitions in Fig.~\ref{fig:spin2q}. We can then prove Eq.~(\ref{eq:Spin2comm}) by a similar argument as the spin-1 case. Specifically, we note that $\ket{T_{3,M}}$, $M = 2,3$ and $\ket{T_{4,M}}$, $M = 3,4$ are the only states with $J=3,4$ present in the scar states $\ket{2\mathcal{S}_{2n}}$. This follows from successive applications of $Q^\dagger$ onto the ground state, which contains states with $J=0,1,2$. More formally, using Fig.~\ref{fig:spin2q}, it is easy to check that the common null space of all $\ketbra{...}{T_{3,M}}_{j,j+1}$, $M < 2$, and $\ketbra{...}{T_{4,M}}_{j,j+1}$, $M < 3$, remains invariant under the action of $Q^\dagger$ [where it is important that $Q^\dagger$ is a sum of on-site terms and the part touching $j$ and $j+1$ is precisely $(-1)^j q_{j,j+1}$].

Evaluating the commutator $\left[H^{(2)}, Q^\dagger \right]$, we can separate it to a $2 Q^\dagger$ term and a term that annihilates the scar states. The below argument formalizes this.

We group the different states as follows~(Fig.~\ref{fig:spin2q}):
\begin{enumerate}[label=(\alph*)]
    \item $\mathcal{G} = \{\ket{T_{J,M}} | J = 0,1,2 \}$ (all two-site spin configurations present in the ground state);
    \item $\mathcal{D} = \{\ket{T_{J,M}} | J = 3, M \geq 2 \text{ or } J = 4, M \geq 3 \}$ (``Dark states" of the system under the action of $q_{j,j+1}$);
    \item $\mathcal{R} = \{\ket{T_{J,M}} | J = 3, M < 2 \text{ or } J = 4, M < 3 \}$ (all spin configurations which are never present in the scar system (``the rest"))~.
\end{enumerate}
Under this grouping, we can write $h_{j,j+1}$ and $q_{j,j+1}$ as:
\begin{align}
    h_{j,j+1} &= \begin{pmatrix}
    \mathbf{0} & \mathbf{0} & \mathbf{0} \\
    \mathbf{0} & \mathbf{I} & \mathbf{0} \\
    \mathbf{0} & \mathbf{0} & \mathbf{I} \\
    \end{pmatrix}_{\mathcal{G,D,R}}~,\label{eq:hmat} \\
    q_{j,j+1} &= \begin{pmatrix}
    \mathbf{0} & \mathbf{0} & q_{\mathcal{G,R}} \\
    q_{\mathcal{D,G}} & \mathbf{0} & q_{\mathcal{D,R}} \\
    \mathbf{0} & \mathbf{0} & q_{\mathcal{R,R}} \label{eq:qmat} \\
    \end{pmatrix}_{\mathcal{G,D,R}} ~,
\end{align}
where each matrix $q_{\mathcal{A,B}}$ encodes the transitions from states in set $\mathcal{B}$ to states in $\mathcal{A}$, with some amplitudes which are irrelevant for the proof. The subscripts $\mathcal{G,D,R}$ indicate the order of the blocks in $h_{j,j+1}$ and $q_{j,j+1}$. The commutator is then simply:
\begin{align}
    \left[h_{j,j+1}, q_{j,j+1} \right] &= \begin{pmatrix}
    \mathbf{0} & \mathbf{0} &  -q_{\mathcal{G,R}} \\
    q_{\mathcal{D,G}} & \mathbf{0} & \mathbf{0} \\
    \mathbf{0} & \mathbf{0} & \mathbf{0} \\
    \end{pmatrix}_{\mathcal{G,D,R}} \nonumber \\
    & = q_{j,j+1} + \begin{pmatrix}
    \mathbf{0} & \mathbf{0} &  -2q_{\mathcal{G,R}} \\
    \mathbf{0} & \mathbf{0} & -q_{\mathcal{D,R}} \\
    \mathbf{0} & \mathbf{0} & -q_{\mathcal{R,R}} \\
    \end{pmatrix}_{\mathcal{G,D,R}}~.
\label{eq:GRDmat}
\end{align}
The second term in Eq.~(\ref{eq:GRDmat}) annihilates the scar states because all states in $\mathcal{R}$ are never present in $\ket{2\mathcal{S}_{2n}}$, as discussed above. Summing over $j$, we get the desired relation Eq.~(\ref{eq:Spin2comm}), noting that $\sum_j (-1)^j q_{j,j+1} = 2 Q^\dagger$.

It is clear from this proof why one could not have arbitrary coefficients for the projectors in the Hamiltonian, e.g.,
$H^{(2)\prime} = \sum_{j=1}^L \left(\alpha P^{(3,2)}_{j,j+1} + \beta P^{(4,2)}_{j,j+1} \right)$. Even though this family of Hamiltonians shares the ground state $\ket{2G}$, there is no scar tower because the commutation relation Eq.~(\ref{eq:Spin2comm}) is not preserved. When $\alpha \neq \beta$, evaluating the commutatator $[H,Q^\dagger]$ will multiply terms $\ketbra{T_{3,M}}{...}$ by $\alpha$ and $\ketbra{T_{4,M}}{...}$ by $\beta$. Summing over $j$, we do not recover the desired $[H,Q^\dagger] = 2Q^\dagger + A$ structure (Eq.~(\ref{eq:comm2})).

On the other hand, the above proof goes through if we replace the identity matrix in the $\mathcal{R},\mathcal{R}$ block in the two-spin Hamiltonian by an arbitrary Hermitian matrix $h_{\mathcal{R,R}}$, which can also vary from site to site.  This then gives generalizations of the spin-2 AKLT model that have the same tower of scar states, similarly to our discussion in the spin-1 case in Sec.~\ref{subsec:familyS1}.

\subsection{Generalization to higher spins}
The generalization to the scar towers in higher-spin 1D AKLT models---also discussed in Ref.~\cite{Moudgalya2018}---follows immediately. The spin-$S$ 1D AKLT model can be written as:
\begin{equation}
    H^{(S)} = \sum_{j=1}^L \sum_{J=S+1}^{2S} P^{(J,S)}_{j,j+1} ~,
\end{equation}
where $P^{(J,S)}_{j,j+1}$ are projectors onto states of total spin $J$, formed by spin-$S$'s on sites $j$ and $j+1$. The tower of scar states is written:
\begin{gather}
    \ket{S\mathcal{S}_{2n}} = (Q^\dagger)^n \ket{SG}, \nonumber \\
    Q^\dagger = \sum_{j=1}^{L} (-1)^j (S^+_j)^{2S}~, ~~~ n=0,...,L/2 ~, \label{eq:Sqj}
\end{gather}
where the ground state $\ket{SG}$ can be similarly expressed through Schwinger bosons~\cite{Moudgalya2018}, and $\ket{S\mathcal{S}_{2n}}$ has energy $E=2n$.

By applying similar selection rules as above, we can group the two-site states into $\mathcal{G}$, $\mathcal{D}$, and $\mathcal{R}$ as follows:
\begin{align}
    \mathcal{G} &= \left\{ \ket{T_{J,M}} | J \leq S \right\} ~, \nonumber \\
    \mathcal{D} &= \left\{ \ket{T_{J,M}} | J > S, M \geq S \right\} ~, \\
    \mathcal{R} &= \left\{ \ket{T_{J,M}} | J > S, M < S \right\} ~. \nonumber
\end{align}
It is easy to verify that states $\mathcal{G}$ are occupied in the ground state, states in $\mathcal{D}$ are subsequently occupied in $\ket{S\mathcal{S}_{2n}}$, and states in $\mathcal{R}$ have 0 overlap with the scar tower.  We note that the grouping here can be made tighter, specifically states $\ket{T_{J,S}}$, for spin $J$ with the same parity as $S$, can be moved from $\mathcal{D}$ to $\mathcal{R}$. The grouping above, however, is sufficient for the proof.

Equations~(\ref{eq:hmat})-(\ref{eq:GRDmat}) are then valid and we obtain the desired commutation relation, and hence the tower of exact scar states.

\subsection{Absence of this scar tower in the 2D AKLT models}
Having established this two-site picture of the exact scar tower, a natural possible extension would be to higher-dimensional AKLT models. We find that these scar states do not appear in those models, and we will illustrate this through the spin-2 AKLT model on a 2D square lattice.

The 2D spin-2 AKLT model consists of spin-2's on a square lattice~\cite{kennedy_two-dimensional_1988}. Similar to other AKLT models, the ground state can be constructed through four Schwinger bosons on each site. One might ask if the same tower of scars exists in 2D, using the same operator $Q^\dagger$ in Eq.~(\ref{eq:2qj}). The crucial difference between the 1D and 2D spin-2 AKLT models is that the 2D Hamiltonian only consists of the projector onto spin-4:
\begin{equation}
    H^{(2)}_{2D} = \sum_{\expval{ij}}  P^{(4,2)}_{\expval{ij}} ~,
\end{equation}
where the sum is taken over all bonds $\expval{ij}$. Because each site has four nearest neighbors (versus two in the 1D chain), the ground state consists of bonds with total spin $J=0,1,2,$ and $3$. Therefore, if we repeat our procedure from above, $q_{\expval{ij}}$ will have terms that send $\mathcal{G}$ to $\mathcal{G}$, specifically $\ketbra{T_{3,2}}{T_{2,-2}}, \ketbra{T_{3,3}}{T_{2,-1}}, \ketbra{T_{2,1}}{T_{3,-3}},$ and $\ketbra{T_{2,2}}{T_{3,-2}}$. These terms violate the important structure in Eqs.~(\ref{eq:qmat}) and (\ref{eq:GRDmat}) and
prevent us from satisfying the relation Eq.~(\ref{eq:Spin2comm}), thus excluding the same scar tower in the 2D case. Similar arguments can be applied to other higher-dimensional AKLT models, such as the spin-3/2 AKLT model on a honeycomb lattice. We note, however, that this does not exclude the possibility of other towers of states (for example related by different $Q^\dagger$ operators) in higher-dimensional AKLT models.

\section{Application to other exact scar towers: the spin-1 XY model}
\label{sec:spin1XY}
We can use the same commutator structure to understand other known exact towers of scar states, namely a perturbed spin-1 XY model studied in Ref.~\cite{schecter_weak_2019}, and a domain-wall-conserving spin-1/2 1D model studied in Ref.~\cite{iadecola_quantum_2019}. 

In Ref.~\cite{schecter_weak_2019}, Schecter and Iadecola introduced an exact tower of scar states in the following perturbed spin-1 XY model on a cubic lattice in arbitrary dimension:
\begin{align}
    H^{IS}_1 &= J\sum_{\expval{ij}} (S_i^x S_j^x + S_i^y S_j^y) + h \sum_j S_j^z + D \sum_j \left(S^z_j \right)^2 \nonumber\\
    &= H_{XY} + H_{z} + H_{z^2} ~.
\end{align}
In 1D, one has to introduce a third-neighbour
term $H_3 = J_3 \sum_j (S_j^x S_{j+3}^x + S_j^y S_{j+3}^y)$ to break a special nonlocal SU(2) symmetry present in sectors with even magnetization~\cite{kitazawa_ansu2_2003,chattopadhyay_quantum_2019}. The scar tower has the same operator $Q^\dagger$ as in the AKLT model, generalized to arbitrary-dimensional cubic lattice of $V$ sites:
\begin{equation}
    \ket{\mathcal{S}^{XY}_n} = \left(Q^\dagger \right)^n \ket{\Omega}, \quad Q^\dagger = \sum_j e^{i \mathbf{r}_j \cdot \pmb{\pi}} \left(S^+_j \right)^2 ~,
    \label{eq:ISXYscar1}
\end{equation}
where $\ket{\Omega}$ is the ferromagnetic state of all `-1's and $n = 0,...,V$. These states were proven in Ref.~\cite{schecter_weak_2019} through a scattering picture, but we can again quickly prove them through a commutator picture. It is immediate that $[H_z, Q^\dagger] = 2h Q^\dagger$. $H_{z^2}$ measures the number of `0's, which is invariant under $Q^\dagger$, so $[H_{z^2}, Q^\dagger] = 0$. We can also quickly compute that
\begin{equation}
    \left[H_{XY}, Q^\dagger \right] = 4J \sum_{\expval{ij}} e^{i \mathbf{r}_i \cdot \pmb{\pi}} \left(\ketbra{0,1}{-1,0} - \ketbra{1,0}{0,-1}\right)_{i,j},
\end{equation}
which annihilates the scar subspace, because all $\ket*{\mathcal{S}^{XY}_n}$ contain no `0's. This proves the scar tower Eq.~(\ref{eq:ISXYscar1}) with scar energies $E_n = h(2n-V) + DV$. The same argument works on any bipartite graph with arbitrary $J_{ij}$.

\subsection{Connection to two-site picture and to Shiraishi-Mori structure}
It is also instructive to consider a more specialized two-site formalism similar to the one developed in the spin-$S$ AKLT model in Sec.~\ref{subsec:proofspin2AKLT}.
Specifically, the $XY$ Hamiltonian on a bond $\expval{ij}$ can be written as
\begin{align*}
h^{XY}_{ij} = \left(\sqrt{2} \ketbra{X_1}{X_2} +  \ketbra{X_3}{X_4} + \ketbra{X_5}{X_6} + \hc \right)_{ij} ~,
\end{align*}
where
\begin{align}
& \ket{X_1} = (\ket{1,-1} + \ket{-1,1})/\sqrt{2}~,~~~ \ket{X_2} = \ket{0,0}, \nonumber\\
& \ket{X_3} = \ket{1,0},~~~ \ket{X_4} = \ket{0,1}, \label{eq:ketXs} \\
& \ket{X_5} = \ket{-1,0},~~~ \ket{X_6} = \ket{0,-1}. \nonumber
\end{align}
We complete the two-site Hilbert space basis with three more states
\begin{align}
& \ket{X_7} = (\ket{1,-1} - \ket{-1,1})/\sqrt{2}~, \\
& \ket{X_8} = \ket{1,1},~~~ \ket{X_9} = \ket{-1,-1}, \nonumber
\end{align}
and write $q_{ij} = (S_i^+)^2 - (S_j^+)^2$ as
\begin{align}
q_{ij} =& 2\left(\ketbra{X_3}{X_5} - \ketbra{X_4}{X_6}\right)_{ij} \\
& + 2\sqrt{2}\left(\ketbra{X_7}{X_9} - \ketbra{X_8}{X_7}\right)_{ij} ~. \nonumber
\end{align}
Crucially, $h^{XY}_{ij}$ contains only states from the set $\mathcal{R} = \{\ket{X_a}|~ a=1,...,6\}$, and hence annihilates everything in the corresponding groups $\mathcal{G} = \{\ket{X_9}\}$ and $\mathcal{D} = \{\ket{X_7},\ket{X_8}\}$, while $q_{ij}$ is block-diagonal with respect to these basis sets. Following the notation of Sec.~\ref{subsec:proofspin2AKLT}, we write:
\begin{align}
    h_{i,j} &= \begin{pmatrix}
    \mathbf{0} & \mathbf{0} & \mathbf{0}\\
    \mathbf{0} & \mathbf{0} & \mathbf{0}\\
    \mathbf{0} & \mathbf{0} & h_{\mathcal{R,R}}\\
    \end{pmatrix}_{\mathcal{G,D,R}}~,\\
    q_{i,j} &= \begin{pmatrix}
    \mathbf{0} & \mathbf{0} & \mathbf{0}\\
    q_{\mathcal{D,G}} & q_{\mathcal{D,D}} & \mathbf{0}\\
    \mathbf{0} & \mathbf{0} & q_{\mathcal{R,R}}\\
    \end{pmatrix}_{\mathcal{G,D,R}}~,
\end{align}
which gives 
\begin{equation}
    \left[h_{i,j},q_{i,j} \right] = \begin{pmatrix}
    \mathbf{0} & \mathbf{0} & \mathbf{0}\\
    \mathbf{0} & \mathbf{0} & \mathbf{0}\\
    \mathbf{0} & \mathbf{0} & \left[h_{\mathcal{R,R}},q_{\mathcal{R,R}}\right]\\
    \end{pmatrix}_{\mathcal{G,D,R}}~.\\
\end{equation}
As with Eq.~(\ref{eq:GRDmat}), this proves the presence of the tower $\ket{\mathcal{S}_n^{XY}}$ and also shows that
\begin{align}
    H' = \sum_{\expval{ij}} \left(h_{ij} \right)_{\mathcal{R}} + h \sum_j S_j^z + D \sum_j \left(S^z_j \right)^2 ~,
\end{align}
defines a family of Hamiltonians that share the scar tower $\ket{\mathcal{S}^n_{XY}}$. Here $\left(h_{ij} \right)_{\mathcal{R}}$ is an arbitrary (possibly site-dependent) Hermitian matrix restricted to the $\mathcal{R}$ subspace. This generalization represents the Shiraishi-Mori embedded Hamiltonian structure~\cite{shiraishi_systematic_2017} known for this scar tower in the spin-1 XY model on any bipartite graph, see Ref.~\cite{schecter_weak_2019}.

Indeed, as long as the graph is connected, the tower $\ket{\mathcal{S}_n^{XY}}$ coincides with the Shiraishi-Mori space defined as the null space of all $\ketbra{...}{X_a}_{ij}, a=1,...,6$.

Given some similarity in such an analysis between the AKLT and the spin-1 XY cases, one may ask what prevented the possibility of recasting the AKLT scars in terms of two-site Shiraishi-Mori projectors.  One difference that we see is that in the AKLT case, there was no separation of states into $\mathcal{R}$ and $\mathcal{R}^C$ sets such that the Hamiltonian is nonzero only in the $\mathcal{R}$ subspace and $Q^\dagger$ is block-diagonal with respect to the two sets.  The presence and structure of the off-diagonal terms in $Q^\dagger$, Eq.~(\ref{eq:qmat}), was actually important and forced some parts of the AKLT Hamiltonian to be fixed for the tower to exist.  Also, while the AKLT tower states $\ket{\mathcal{S}_{2n}}$ lie in the common null space of projectors constructed from the spin-1 AKLT ``$\mathcal{R}$" states $\{\ket{T_{2,M}}_{j,j+1},~M = -2,1,0\}$, we numerically observe additional states in this null space. While we see these specific differences, it is fair to say that at present we do not know general rules that would allow us to see why the AKLT tower cannot be realized in the embedded Hamiltonian approach.

\subsection{Additional scar tower in 1D}
In 1D, with $J_3 = D = 0$, Schecter and Iadecola also numerically observed and conjectured the following scar tower:
\begin{equation}
    \ket{\mathcal{S}^{XY,2}_n} = \sum_{i_1 \neq \cdots \neq i_n} \!\!\!\! (-1)^{\sum_j i_j} (S_{i_1}^+ S_{i_1+1}^+) \cdots (S_{i_n}^+ S_{i_n+1}^+) \ket{\Omega}~,
    \label{eq:ISXYscar2}
\end{equation}
with energies $E'_n = h(2n-L)$.  These states were subsequently proven by Chattopadhyay et.~al.~\cite{chattopadhyay_quantum_2019} by compressing all the scar states into a single MPS state, which they then proved is an eigenstate of $H_{XY}$. The term $V = \epsilon \sum_j (S^+_j)^2 (S^-_{j+1})^2 + \hc$ was also added to the Hamiltonian to destroy integrability (and with this term the previous tower is no longer exact).

While the operator in Eq.~(\ref{eq:ISXYscar2}) cannot be cast into the ladder form $(Q^\dagger)^n$, we can still prove that they are eigenstates directly through $H_{XY} \ket*{\mathcal{S}^{XY,2}_n} = 0$. This is done in Appendix~\ref{sec:XYProof}.

\section{Common ``parent" Hamiltonians for the spin-1 AKLT and XY model scars}
\label{sec:parentham}
In this section, we discuss a common underlying Hamiltonian $H_0$ that hosts the scar towers from both the spin-1 AKLT and XY models. While the two models are different at face value, both scar towers involve the common operator $Q^\dagger = \sum_j (-1)^j (S_j^+)^2$. The underlying model possesses the symmetry $\left[H_0, Q^\dagger \right] = 0$, which produces both scar towers. In fact, $H_0$ has SU(2) symmetry (unrelated to the spin-1 SU(2) symmetry in the original AKLT model), and we lastly discuss a new nonintegrable model that contains both towers of states.

We first define
\begin{equation}
    H_0 = \sum_{\expval{ij}} \left(\ketbra{1,0}{0,1} - \ketbra{-1,0}{0,-1} + \hc \right)_{i,j} ~,
\end{equation}
where we can take any bipartite graph in the XY model scar case, while we specialize to 1D chain in the AKLT case. We can immediately verify that $H_0$ commutes with $Q^\dagger$ as defined in Eq.~(\ref{eq:ISXYscar1}).
As noted in Ref.~\cite{schecter_weak_2019}, operators $J^+ \equiv Q^\dagger/2$, its Hermitian conjugate $J^- \equiv Q/2$, and $J^z = (1/2) \sum_j S^z_j$ form a standard $su(2)$ algebra, $[J^z, J^\pm] = \pm J^\pm$, $[J^+, J^-] = 2 J^z$. Hence, the Hamiltonian $H_0$ has a global SU(2) ``pseudospin" symmetry. In 1D, $H_0$ is in fact integrable and can be transformed into a hopping problem of spin-1/2 hard core bosons with no double occupancy, as discussed in Appendix~\ref{sec:H0appendix}.

The Hamiltonian $H_0$ is significant for the scar tower $\ket{\mathcal{S}_n^{XY}}$ in the spin-1 XY model, as follows. We rewrite $H_{XY}$ as
\begin{equation}
    H_{XY} = H_0 + H'_{XY} ~,
\end{equation}
where, using the notation in Eq.~(\ref{eq:ketXs}),
\begin{equation}
    H'_{XY} = \sum_j \big(\sqrt{2}\ketbra{X_1}{X_2} + 2\ketbra{X_5}{X_6} + \hc \big)_{j,j+1} ~.
\end{equation}
Given that $H_0$ annihilates the ferromagnetic state $\ket{\Omega}$, applying the ladder operator $Q^\dagger$ produces a set of zero-energy eigenstates $\{\ket{\mathcal{S}_n^{XY}}\}$ in Eq.~(\ref{eq:ISXYscar1}).
These states span the total pseudospin $J = L/2$ sector, the largest-pseudospin multiplet formed by $L$ ``spin-1/2" objects. It can be easily verified that $H'_{XY}$ annihilates $\ket{\mathcal{S}_n^{XY}}$, so the exact eigenstates persist under addition of $H'_{XY}$, giving the scar tower in $H_{XY}$. 

$H_0$ also hosts the scar tower of the AKLT model. We first notice that
\begin{gather}
    \frac{1}{2} H_0 + \sum_j \frac{1}{2}(S^z_j + S^z_{j+1}) = 
    \sum_j \big(\ketbra{T_{2,2}}{T_{2,2}} \label{eq:H0plusSztot} \\
    + \ketbra{T_{2,1}}{T_{2,1}} - \ketbra{T_{2,-1}}{T_{2,-1}} - \ketbra{T_{2,-2}}{T_{2,-2}} \big)_{j,j+1} ~. \nonumber
\end{gather}
This allows us to express $H_{AKLT}$ as
\begin{equation*}
    H_{AKLT} = \frac{1}{2} H_0 + H'_{AKLT} + \sum_j \frac{1}{2}(S^z_j + S^z_{j+1}) ~,
\end{equation*}
where
\begin{align}
    H'_{AKLT} &= \sum_j \big(2\ketbra{T_{2,-2}}{T_{2,-2}} + 2\ketbra{T_{2,-1}}{T_{2,-1}} \\ 
    & + \ketbra{T_{2,0}}{T_{2,0}} \big)_{j,j+1} ~. \nonumber
\end{align}
From our discussion in Section~\ref{sec:spin1AKLT}, $H'_{AKLT}$ annihilates the scar states $\ket{\mathcal{S}_{2n}}$. In PBC, the term $\sum_j \frac{1}{2}(S^z_j + S^z_{j+1}) = \sum_j S^z_j$, with commutation $\left[\sum_j S^z_j, Q^\dagger \right] = 2 Q^\dagger$~\footnote{In OBC, $[\sum_j \frac{1}{2}(S^z_j + S^z_{j+1}), Q^\dagger] = 2Q^\dagger + (S^+_1)^2 - (S^+_L)^2$. The last two terms annihilate the scar states in OBC with both ``dangling'' spin-1/2's pointing up~\cite{Moudgalya2018, moudgalya_entanglement_2018}, which can be combined with the commutator argument to prove these scar states in OBC}.

The AKLT ground state $\ket{G}$ is annihilated by $H_0$. This is most easily seen in PBC through Eq.~(\ref{eq:H0plusSztot}) and the fact that $\sum_j S^z_j = 0$ in $\ket{G}$~\footnote{This fact can also be seen directly through the following argument. By the string order of the AKLT ground state, `-1's must be followed by `1's and vice versa, with any number of `0's in between. $H_0$ preserves this string order. To show that the image $H_0 \ket{G} = 0$, we consider any product state in $H_0 \ket{G}$. This has string order and therefore has equal number of `1,0' and `0,-1' substrings, because they occur in substrings `1,0,...,0,-1'. Therefore $n_{(1,0)} = n_{(0,-1)}$ and likewise $n_{(-1,0)} = n_{(0,1)}$. Any given product state has $n_{(1,0)} + n_{(0,1)}$ preimage product states under $H_0$, in which a `1' was hopped, and $n_{(-1,0)} + n_{(0,-1)} = n_{(0,1)} + n_{(1,0)}$ preimages in which a `-1' was hopped. From the MPS in Eq.~(\ref{eq:MPS}), using $A^{[\pm1]} A^{[0]} = -A^{[0]} A^{[\pm1]}$, all preimages have the same coefficient in $\ket{G}$. These preimages cancel in $H_0\ket{G}$ because under $H_0$, hopping a `-1' has negative sign. Therefore $H_0 \ket{G} = 0$.}. By applying $Q^\dagger$, we generate a set of zero-energy eigenstates $\ket{\mathcal{S}_n}$ of $H_0$. These states persist under the perturbation $H'_{AKLT}$. Since these states have well-defined $J^z$ and the perturbation terms commute with $J^z$, adding the operator $\sum_j S^z_j = 2J^z$ generates the equal energy spacing of $\ket{\mathcal{S}_n}$ seen in the AKLT model.

We note that the Hamiltonian $H_0$ that commutes with $Q^\dagger$ [and hence has the discussed SU(2) symmetry] and that annihilates the AKLT scar tower is not unique.  Specifically, from the considerations in Sec.~\ref{sec:spin1AKLT} it is easy to check that $\sum_j c_j \ketbra{T_{2,0}}{T_{2,0}}_{j,j+1}$ with arbitrary $c_j$ also has these properties and can be added to $H_0$. The spin-1 XY model $\pi$-bimagnon tower is automatically in the exact spectrum based solely on the fact that it represents the highest $J$ multiplet under this SU(2) (in the specific model, the energy is zero since each $\braket{T_{2,0}}{\Omega}_{j,j+1} = 0$).  In this way, in this model the $\pi$-bimagnon tower of Iadecola and Schecter are like the $\eta$-pairing states in the Hubbard model on bipartite lattices; this tower becomes true scars once this SU(2) symmetry is broken while preserving the tower as exact eigenstates, as happens in the spin-1 XY model and its generalizations.

On the other hand, the appearance of the AKLT tower is nontrivial, since the AKLT state does not have definite total $J$ quantum number.  In particular, each total pseudospin $J$ component of the AKLT state is also a zero-energy eigenstate of the Hamiltonian. That is, $H P_{J=n}\ket{G} = 0$, where $P_{J=n}$ is the projector onto the sector of total pseudospin $n$. One can prove that all such pseudospin components are nonzero, except for $J=L/2$ at $L \equiv 2 ~\text{(mod 4)}$~\footnote{To prove this, consider the following maximally spin-polarized $J=n$ states: $\ket*{1, \dots, 1_{2n}, 0, \dots, 0}$ for even $n$, and $\ket*{1, 0, 1, \dots, 1_{2n+1}, 0, \dots, 0}$ for odd $n$. Applying $(J^-)^n$, we obtain states with nonzero overlap with the string-ordered AKLT ground state, namely with the AKLT contributions $\ket*{1, -1, \dots, 1, -1_{2n}, 0, \dots, 0} + \ket*{-1, 1, \dots, -1, 1_{2n}, 0, \dots, 0}$ for even $n$ and $-\ket*{1, 0, -1, \dots, 1, -1_{2n+1}, 0, \dots, 0} - \ket*{-1, 0, 1, \dots, -1, 1_{2n+1}, 0, \dots, 0}$ for odd $n$. This construction works for all $n < L/2$ (and for $n=L/2$ when $L/2$ is even).}. This exception is related to our remark in Sec.~\ref{sec:spin1AKLTexactscarstates} that when $L \equiv 2 ~\text{(mod 4)}$, the ferromagnetic state $\ket{1,1,\dots,1}$ does not belong to the AKLT scar tower. We note, however, that each pseudospin component $P_{J=n} \ket{G}$ does not preserve properties of $\ket{G}$ such as the ``string-order", and understanding the properties of these states could be interesting future work.

Exact diagonalization on the model $H_0 + \sum_j \ketbra{T_{2,0}}{T_{2,0}}_{j,j+1}$ for $L \leq 12$ confirms that $P_{J=n} \ket{G}$ are zero-energy eigenstates in otherwise nonintegrable spectra at sectors $J^z=0, k=0, I=1, J=n$, and other sectors related by $J^{\pm}$. These states also appear to be bipartite entanglement entropy outliers, confirming that the pseudospin $J=n$ components of the AKLT scar towers remain scars in this nonintegrable model.

\section{The domain-wall-conserving spin-1/2 model}
\label{sec:spin1/2model}
In Ref.~\cite{iadecola_quantum_2019}, Iadecola and Schecter also introduced an exact tower of scar states in the following model on a 1D spin-1/2 chain of $L$ sites:
\begin{align}
    H_{1/2}^{IS} &= \sum_{j=1}^L \lambda \left(\sigma^x_j - \sigma^z_{j-1} \sigma^x_j \sigma^z_{j+1} \right) + \Delta \sigma^z_j + J \sigma^z_j \sigma^z_{j+1} \nonumber\\
    &= H_\lambda + H_z + H_{zz} ~,
    \label{eq:HIS1/2}
\end{align}
where $\sigma^x,~\sigma^z$ are the $x$ and $z$ Pauli matrices. The $H_\lambda$ term describes moving of domain walls `01' and `10' (identifying $\ket{\uparrow} \equiv \ket{1}$, $\ket{\downarrow} \equiv \ket{0}$). $H_z$ and $H_{zz}$ are field operators that measure the magnetization and number of domain walls respectively. Consequently the number of domain walls $N_{DW} = \sum_j (1 - \sigma^z_j \sigma^z_{j+1})/2 = \sum_j (\ketbra{01}{01} + \ketbra{10}{10})_{j,j+1}$ is a conserved quantity in this model. (This conservation law is somewhat unusual in that it is not realized as an onsite symmetry in the model.) On the other hand, the magnetization $\sum_j \sigma^z_j$ is not conserved by the Hamiltonian, but plays an important role for the scar states described below.

This model also exhibits particle-hole symmetry in each symmetry sector: $\prod_j \sigma^y_j (H_\lambda + H_z + H_{zz}) \prod_j \sigma^y_j = - H_\lambda - H_z + H_{zz}$, and $H_{zz}$ is simply a constant on every symmetry sector, because $N_{DW}$ is fixed. Therefore in each sector, the spectrum is symmetric about $E = J(L - 2 N_{DW})$~\footnote{We numerically observe a large number of degenerate states at the symmetry point $E = J(L - 2 N_{DW})$ of each sector.}.

This model has also recently been studied in Ref.~\cite{borla_confined_2019} as a $\mathbb{Z}_2$ lattice gauge theory coupled to spinless fermions, and in Ref.~\cite{yang_hilbert-space_2019} in the context of Hilbert space fragmentation in the limit $\Delta \gg \lambda$. We note that the model Eq.~(\ref{eq:HIS1/2}) at general parameters does not exhibit Hilbert space fragmentation. This is evident from the fact that all real-space configurations with fixed $N_{DW}$ can be connected to the common string $\ket*{\underbracket{1010\cdots10}_{N_{DW}}0\cdots0}$ via repeated applications of $H_\lambda$. Reference~\cite{ostmann_localization_2019} also studied this model with disorder and suggested an experimental realization through ``antiblockaded" Rydberg atomic lattices.

The scar tower in Ref.~\cite{iadecola_quantum_2019} is written (up to normalization):
\begin{equation}
    \ket*{\mathcal{S}^{IS}_n} = (R^\dagger)^n \ket{\Omega},~~ R^\dagger = \sum_{j=1}^L (-1)^j P^0_{j-1} \sigma_j^+ P^0_{j+1} ~,
\end{equation}
where $\ket{\Omega} = \ket{00\cdots0}$ is the ferromagnetic state with all spins down, $P^0_j = \ketbra{0}{0}_j$ is the projector onto the down state, and $\sigma^+_j = \ketbra{1}{0}_j$. The scar states $\ket*{\mathcal{S}^{IS}_n}$ have energy $E_n = (2\Delta - 4J) n + (J - \Delta) L$ in PBC. (In OBC the energy is reduced by $J$.) They are equally spaced in energy due to the field operators $H_z$ and $H_{zz}$. It is also notable that because of the projectors in $R^\dagger$, these states obey the Rydberg constraint that no two `1's are adjacent. Therefore the scar tower terminates at $n=L/2$ with the $\mathbb{Z}_2$ state $\left(\ket{1 0 \cdots 1 0} +(-1)^{L/2} \ket{0 1 \cdots 0 1}\right)/\sqrt{2}$. This Rydberg constraint is nowhere present in the Hamiltonian and is therefore dubbed an ``emergent kinetic constraint" by Iadecola and Schecter.

The scar states were originally proven through cancellation of scattering terms, but we can also prove this through the commutator picture. We can easily show that $\ket{\Omega}$ is annihilated by $H_\lambda$ and is trivially an eigenstate of $H_z$ and $H_{zz}$. Therefore, proving the tower of states amounts to showing that:
\begin{equation}
    \left[H_{1/2}^{IS}, R^\dagger \right] \ket{\mathcal{S}^{IS}_n} = (2\Delta - 4J) R^\dagger \ket{\mathcal{S}^{IS}_n} ~.
    \label{eq:ISHcomm1}
\end{equation}
A quick calculation yields:
\begin{equation}
    \left[H_z, R^\dagger \right] = 2\Delta R^\dagger~~\text{and}~~\left[H_{zz}, R^\dagger \right] = -4J R^\dagger ~.
\end{equation}
Computing the commutator with $H_\lambda$ is also straightforward but shows nontrivial structure:
\begin{align}
   \label{eq:Hlambdacomm}
    &\left[H_\lambda, R^\dagger \right] \\
    &= 2\lambda \sum_{j=1}^L (-1)^j \left(P^{1}_{j-1} \sigma^-_j \sigma^+_{j+1} P^{0}_{j+2} - P^{0}_{j-1} \sigma^+_j \sigma^-_{j+1} P^{1}_{j+2} \right) ,\nonumber
\end{align}
where $P^{1}_j = \ketbra{1}{1}_j$ is the projector onto the up spin, and $\sigma^-_j = \ketbra{0}{1}_j$. Each local term in Eq.~(\ref{eq:Hlambdacomm}) annihilates the states $\ket*{\mathcal{S}^{IS}_n}$, because $P^{1}_{j-1} \sigma^-_j$ and $\sigma^-_{j+1} P^{1}_{j+2}$ are only nonzero on $\ket{11}$ on the respective sites, which is disallowed by the Rydberg constraint. Therefore $\left[H_\lambda, R^\dagger \right] \ket*{\mathcal{S}^{IS}_n} = 0$, proving the desired Eq.~(\ref{eq:ISHcomm1}).

Iadecola and Schecter also obtained a conjugated scar tower through the global $\mathbb{Z}_2$ transformation $G = \prod_j \sigma^x_j$, which globally exchanges all `0's and `1's: $\ket{\mathcal{S}'^{IS}_n} = G\ket{\mathcal{S}^{IS}_n}$, with energies $E'_n = -(2\Delta + 4J)n + (J + \Delta)L$. These satisfy
\begin{equation}
 \ket{\mathcal{S}'^{IS}_n} = (R'^\dagger)^n \ket{\Omega'} ~,
 \label{eq:conjscars}
\end{equation}
where $R'^\dagger = G R^\dagger G = \sum_{j=1}^L (-1)^j P^1_{j-1} \sigma_j^- P^1_{j+1}$ and $\ket{\Omega'} = \ket{11\cdots1}$. We also note that on states of fixed magnetization $\sum_j \sigma^z_j$, the action of $G$ is equivalent to that of the particle-hole symmetry $\prod_j \sigma^y_j$, up to a sign factor. Here and below we use $G$ for conceptual simplicity. 

\subsection{New ``pyramid" of exact states in the spin-1/2 model}
\label{sec:orthogonaltowers}
Here we introduce a new set of towers of exact states in the Iadecola-Schecter spin-1/2 model, with PBC and $L$ even.  These towers are organized in a structure which we dub a ``pyramid." We found these states originally in our exact diagonalization numerical studies.
They are, for all $1 \leq n \leq L/2-1$:
\begin{equation}
    \ket{\mathcal{S}^{\text{pyr.}}_{n,m}} = (\mathcal{P}^\dagger)^m \ket{\mathcal{S}^{IS}_n},~~~ m = 0,1,...,L-2n ~, \label{eq:orthogonalscars}
\end{equation}
where
\begin{align}
    \mathcal{P}^\dagger &=\sum_{j=1}^L \sum_{l=1}^{L-2} P^{\mathbf{1},l}_{j-1} \sigma_j^+ P^0_{j+1} \\
    &= \sum_{j=1}^L P^{1}_{j-1} \sigma_j^+ P^0_{j+1} + P^{1}_{j-2} P^{1}_{j-1} \sigma_j^+ P^0_{j+1} + ... ~, \nonumber
\end{align}
and
\begin{equation}
    P^{\mathbf{1},l}_{j-1} = \ketbra*{\underbracket{1\cdots1}_l}{\underbracket{1\cdots1}_l}_{j-l,...,j-1} ~.
\end{equation}
Here $\mathcal{P}^\dagger$ is a nonlocal operator, which enlarges domains of $l$ `1's by one unit to the right, with coefficient $l$.  Note that such a move is allowed only if there are at least two `0's separating the domain being enlarged from the next domain of `1's to the right, i.e., the move is not allowed to merge domains of `1's.

As with $\ket{\mathcal{S}^{IS}_{n}}$, these states have $N_{DW} = 2n$ and $k = n\pi$ (mod $2\pi$), these quantities being unchanged by $\mathcal{P}^\dagger$. The bond inversion number is $I_b = (-1)^{n+m}$ (this fact will become clear from the wavefunction picture in Section~\ref{sec:characterizingSnm}).
$[H_z, \mathcal{P}^\dagger] = 2\Delta \mathcal{P}^\dagger$, so the magnetization increases by 2 and these states have energies $E_{n,m} = E_n + 2\Delta m = 2\Delta (n+m) - 4J n + (J - \Delta)L$. We emphasize here that the magnetization is not a conserved quantity in the Hamiltonian, but is well defined for all $\ket{\mathcal{S}^{\text{pyr.}}_{n,m}}$. 

\begin{figure}
    \centering
\includegraphics[width=0.5\textwidth]{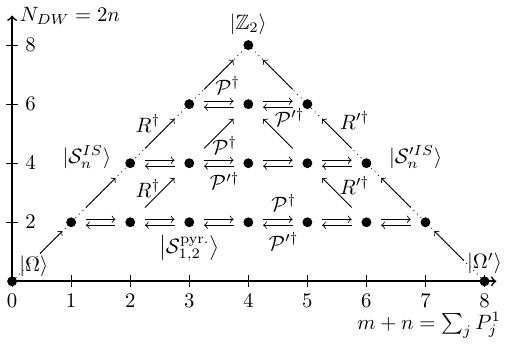}
\caption{``Pyramid" of exact states $\ket{\mathcal{S}^{\text{pyr.}}_{n,m}}$, Eq.~(\ref{eq:orthogonalscars}), for system size $L=8$. Each point represents an exact state, with $n$ increasing upwards ($n = N_{DW}/2$) and $m$ increasing rightwards ($m = \sum_j P^1_j - n$). The vertical axis is the (conserved) number of domain walls $N_{DW}$.  The horizontal axis is the total number of `1's: $\sum_j P^1_j$, which is not a conserved quantity under the Hamiltonian, but these states are eigenstates of this operator with eigenvalue $n+m$. Ladder operators $R^\dagger$, $R'^\dagger$, $\mathcal{P}^\dagger$, and $\mathcal{P}'^\dagger$ are labeled on their respective towers. Ferromagnetic states $\ket{\Omega}$, $\ket{\Omega'}$, CDW state $\ket{\mathbb{Z}_2}$, and a sample state $\ket{\mathcal{S}^{\text{pyr.}}_{1,2}}$ are also labeled. The Iadecola-Schecter towers are the slopes marked with a dotted line, and our new towers are the horizontal lines.
}
\label{fig:towersofstates}
\end{figure}

While the operator $R^\dagger$ increases $N_{DW}$ by 2 in the original Iadecola-Schecter towers, these new towers lie in sectors of constant $N_{DW}$. We also remark that for fixed $n$, each tower $\ket{\mathcal{S}^\text{pyr.}_{n,m}}$ starts from the Iadecola-Schecter scars $\ket{\mathcal{S}^{IS}_{n}}$ and ends at their conjugated scars $\ket*{\mathcal{S}'^{IS}_{n}}$. The various towers form a pyramid-like structure, with the Iadecola-Schecter towers as the outer slopes and our new towers forming the horizontal levels.  This is illustrated in Fig.~\ref{fig:towersofstates}. It is fairly easy to prove these states in Sec.~\ref{sec:directprooforthogonal} once we further characterize the states $\ket{\mathcal{S}^{\text{pyr.}}_{n,m}}$ in Sec.~\ref{sec:characterizingSnm}. We also give an alternate proof in Appendix~\ref{sec:prooftoorthogonal} by showing that $[H_\lambda, \mathcal{P}^\dagger] \ket{\mathcal{S}^{\text{pyr.}}_{n,m}} = 0$.

Taking the global $\mathbb{Z}_2$ transformation $G = \prod_j \sigma^x_j$, we can define the conjugate operator $\mathcal{P}'^\dagger = G \mathcal{P}^\dagger G$ such that $\mathcal{P}'^\dagger \ket{\mathcal{S}^{\text{pyr.}}_{n,m}} = \ket{\mathcal{S}^{\text{pyr.}}_{n,m-1}}$ (up to a numerical factor). $\mathcal{P}'^\dagger$ grows domains of `0's to the right and seemingly does not undo $\mathcal{P}^\dagger$. However, we note that we could have defined $\mathcal{P}^\dagger$ as growing domains to the left, and this would have produced equivalent states (up to a sign factor),
which will become particularly clear after Sec.~\ref{sec:characterizingSnm}.
Therefore, we can also go from $\ket{\mathcal{S}^{\text{pyr.}}_{n,m}}$ to $\ket{\mathcal{S}^{\text{pyr.}}_{n,m-1}}$ by growing domains of `0's either to the left or right.

\begin{figure}
    \centering
    \includegraphics[width=0.5\textwidth]{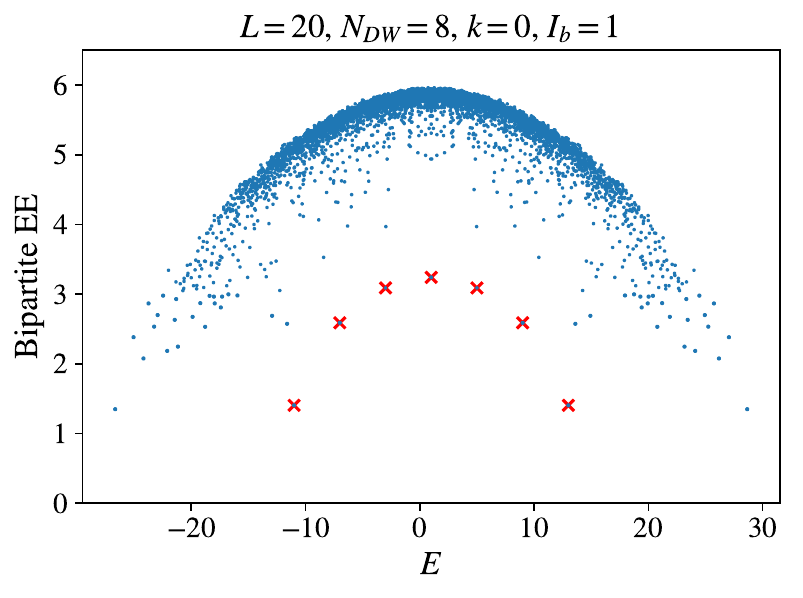}
    \caption{Plot of bipartite EE computed in a $L=20$ chain with PBC, for $\lambda=1, \Delta=1, J=0.25$. We display the $N_{DW}=8, k=0, I_b=1$ sector, with Hilbert space dimension 6478. The pyramid states $\ket{\mathcal{S}_{4,m}^\text{pyr.}}$ for $m=0,2,4,6,8,10,12$ are marked with red crosses. The spectrum also clearly exhibits particle-hole symmetry about $E = J(L - 2 N_{DW}) = 1$.}
    \label{fig:bipEE}
\end{figure}

Lastly, we note that numerical calculations of bipartite entanglement entropy (EE) reveal these states to be EE outliers, as illustrated in Fig.~\ref{fig:bipEE}. While this is not a proof of sub-volume law EE scaling, this suggests that these ``pyramid" states are indeed scar states~\footnote{We note that we do not find the exact pyramid states in ED with OBC (their proof also requires cancellations that do not happen in OBC).  However, we expect that dynamical signatures of the pyramid states (such as behavior of local observables under characteristic quenches) will be the same in large systems irrespective of the boundary conditions}.

\subsection{Characterization of $\ket{\mathcal{S}_{n,m}^\text{pyr.}}$}
\label{sec:characterizingSnm}
It is instructive to discuss these states for small $n$ (at the base of the ``pyramid"). For $n=1$, these states are the set of ``bound magnons":
\begin{equation}
    \ket{\mathcal{S}_{1,m}^\text{pyr.}} = \sum_{j=1}^L (-1)^j \ket*{0\cdots 0\underbracket{1_j1\cdots1}_{m+1}0\cdots0}, 
    \label{eq:Bm}
\end{equation}
for $1 \leq m \leq L-2$ (here and when writing other $\ket{\mathcal{S}_{n,m}^\text{pyr.}}$ states below, we drop overall numerical factors). One can verify easily that these are annihilated by $H_\lambda$. These states comprise the whole $N_{DW} = 2,~k = \pi,~I = \pm 1$ sectors. The other $(k, I)$ sectors with $N_{DW} = 2$ are in fact also exactly solvable and are discussed in Appendix~\ref{sec:NDW2}.

For $n=2$, we discuss the cases $m=1,2$. $m=1$ is the state:
\begin{align}
    \ket{\mathcal{S}_{2,1}^\text{pyr.}} = \sum_{i < j} (-1)^{i+j}& \Big(\ket{\cdots01_i0\cdots 01_j10 \cdots} \\
    &+ \ket{\cdots01_i10\cdots 01_j0 \cdots} \Big)~, \nonumber
\end{align}
with the understanding that only the states with two separated domains of `1's are allowed. In this case, one of the domains must be `1' and the other must be `11'; the domains can be anywhere on the chain subject to not touching; and each such configuration contributes with the specific sign.

Larger $m$'s show more nontrivial patterns of domains. For example, for $m=2$:
\begin{align}
    &\ket{\mathcal{S}_{2,2}^\text{pyr.}} = \sum_{i < j} (-1)^{i+j} \Big(2 \ket{\cdots01_i0\cdots 01_j110 \cdots} \\
    &+2\ket{\cdots01_i11\cdots 01_j0 \cdots} + 2\ket{\cdots01_i10\cdots 01_j10 \cdots}\Big) ~, \nonumber
\end{align}
again with the understanding that only the states with nontouching domains are allowed. In this case, the two domains can be `1'/`111' or `11'/`11' type, but again all configurations have the same weight and specific signs. The $\ket{\cdots01_i10\cdots 01_j10 \cdots}$ term gains a factor of two because it has two preimages in $\ket{\mathcal{S}^\text{pyr.}_{2,1}}$, with the `11' at the $i$ site or $j$ site, while the $\ket{\cdots01_i0\cdots 01_j110 \cdots}$-type terms gain an equal factor of two from $\mathcal{P}^\dagger$. Similarly, $\ket{\mathcal{S}^\text{pyr.}_{2,3}}$ contains an equal superposition of states with `1'/`1111' and `11'/`111' domains. 

This characteristic holds generally, and we can write $\ket*{\mathcal{S}^\text{pyr.}_{n,m}}$ as the superposition of domains of `1's, each with wavenumber $k=\pi$:
\begin{equation}
    \ket{\mathcal{S}^\text{pyr.}_{n,m}} =
    \sum_{i_1 < i_2 < ... < i_n} \sum_{(l_j)} (-1)^{\sum_j i_j} \ket{i_1,...,i_n}_{(l_1,...,l_n)} ~,
    \label{eq:Snmexpansion}
\end{equation}
where $\{i_j\}$ denotes the starting positions of the $n$ domains with lengths $l_1,...,l_n$, constrained such that the domains are at least 1 site apart. The sums are over all possible $\{i_j\}$ and all domain lengths $(l_j)$ obeying $\sum_j l_j = n + m$. 
Thus, $\ket{\mathcal{S}^\text{pyr.}_{n,m}}$ is an equal weight superposition of all states with $N_{DW} = 2n$ and $\sum_j P^1_j = n + m$, with each domain carrying wavenumber $k=\pi$ --- taking reference from the start points of all domains. We can also take reference from all domain end points, or describe $\ket{\mathcal{S}^\text{pyr.}_{n,m}}$ as a superposition of all domains of `0's with $\sum_j P^0_j = L - m - n$ and take reference from their start or end points. These descriptions are all equivalent up to sign factors.

We will now see that the nonlocality of $\mathcal{P}^\dagger$ is in fact necessary to construct this equal superposition, by compensating for combinatorial factors. To obtain Eq.~(\ref{eq:Snmexpansion}) from Eq.~(\ref{eq:orthogonalscars}), note that a given configuration $\ket{i_1,...,i_n}_{(l_1,...,l_n)}$ of $n$ domains with lengths $(l_1,...,l_n)$ must come from the state $\ket{i_1,...,i_n}_{(1,...,1)}$ in $\ket{\mathcal{S}^{IS}_n}$, because domains grow to the right. We express the action of $(\mathcal{P}^\dagger)^m$ on this state as:
\begin{equation*}
  (\mathcal{P}^\dagger)^m \ket{i_1,...,i_n}_{(1,..,1)} = (p^\dagger_{i_1} + ... +p^\dagger_{i_n})^m \ket{i_1,...,i_n}_{(1,..,1)}, 
\end{equation*}
where $p^\dagger_{i_j}$ grows the domain starting at site $i_j$, of length $l$, with coefficient $l$. Therefore when a domain grows from length 1 to $l_j$ it accumulates a factor of $(l_j-1)!$. The term in the expansion of $(p^\dagger_{i_1} + ... + p^\dagger_{i_n})^m$ which produces the state $\ket{i_1,...,i_n}_{(l_1,...,l_n)}$ is $(p^\dagger_{i_1})^{l_1-1}\cdots (p^\dagger_{i_n})^{l_n-1}$. From the multinomial expansion, this has coefficient $m!/((l_1-1)!\cdots(l_n-1)!)$. The operators $(p^\dagger_{i_1})^{l_1-1}\cdots (p^\dagger_{i_n})^{l_n-1}$ compensate this with a coefficient $((l_1-1)!\cdots(l_n-1)!)$. Therefore every state $\ket{i_1,...,i_n}_{(l_1,...,l_n)}$ in  $\ket*{\mathcal{S}^\text{pyr.}_{n,m}}$ is multiplied by the same factor $m!$, giving the desired Eq.~(\ref{eq:Snmexpansion}).

\subsection{Proof of $\ket{\mathcal{S}_{n,m}^\text{pyr.}}$}
\label{sec:directprooforthogonal}
Having established the allowed real-space configurations and their amplitudes in $\ket{\mathcal{S}_{n,m}^\text{pyr.}}$, we can immediately prove that $H_\lambda \ket{\mathcal{S}_{n,m}^\text{pyr.}} = 0$. To do so we note the action of $H_\lambda$ is to grow domains of `1's to the left or right, when possible, and to shrink domains of `1's to the left or right, whenever that domain has length $l>1$:
\begin{align}
& \sigma_j^x - \sigma_{j-1}^z \sigma_j^x \sigma_{j+1}^z = 2(P_{j-1}^1 \sigma_j^x P_{j+1}^0 + P_{j-1}^0 \sigma_j^x P_{j+1}^1) \\
& = 2(\ketbra{110}{100} + \ketbra{100}{110} + \ketbra{011}{001} + \ketbra{001}{011}) ~. \nonumber
\end{align}
The shrinking of domains of `1's can be equivalently thought of as growing domains of `0's to the left or right, whenever possible. We accordingly split $H_\lambda = H_\lambda^{(1)} + H_\lambda^{(0)}$, with $H_\lambda^{(1)}$ the terms that grow domains of `1's and $H_\lambda^{(0)}$ the terms that grow domains of `0's.

To show that $H_\lambda^{(1)}\ket{\mathcal{S}_{n,m}^\text{pyr.}} = 0$, we note that given a configuration of domains in the image $H_\lambda^{(1)}\ket{\mathcal{S}_{n,m}^\text{pyr.}}$
\begin{equation*}
    \ket*{\cdots0\underbracket{1_{i_j}1\cdots1}_{l}0\cdots}~,
\end{equation*}
for each domain of length $l > 1$ like the exhibited $j$th domain above, has two preimages in $\ket{\mathcal{S}_{n,m}^\text{pyr.}}$ in which the $j$th domain was grown:
\begin{equation*}
    (-1)^{i_j} \left(\ket*{\cdots0\underbracket{1_{i_j}1\cdots1}_{l-1}00\cdots} - \ket*{\cdots00\underbracket{1_{i_j+1}1\cdots1}_{l-1}0\cdots}\right)~.
\end{equation*}
The two preimages have opposite signs because the corresponding domain is shifted by one between the two, and hence the corresponding contributions in $H_\lambda^{(1)}\ket{\mathcal{S}_{n,m}^\text{pyr.}}$ cancel. This is true for every domain in $H_\lambda^{(1)}\ket{\mathcal{S}_{n,m}^\text{pyr.}}$, except for those with length $l=1$, for which there is no preimage to consider because that domain could not have been grown. Since all configurations in $H_\lambda^{(1)}\ket{\mathcal{S}_{n,m}^\text{pyr.}}$ have at least one domain of length $l>1$, it follows that $H_\lambda^{(1)}\ket{\mathcal{S}_{n,m}^\text{pyr.}} = 0$.

Since the descriptions of $\ket{\mathcal{S}_{n,m}^\text{pyr.}}$ are equivalent for domains of `0's, this also shows that $H_\lambda^{(0)}\ket{\mathcal{S}_{n,m}^\text{pyr.}} = 0$. This proves the desired $H_\lambda \ket{\mathcal{S}_{n,m}^\text{pyr.}} = 0$ and therefore that $\ket{\mathcal{S}_{n,m}^\text{pyr.}}$ are eigenstates of $H^{IS}_{1/2}$. The ``equal weight superposition" of all domain length configurations was essential to let us move from the description of domains of `1's to domains of `0's. An alternate proof is given in Appendix~\ref{sec:prooftoorthogonal}, by showing that $[H_\lambda, \mathcal{P}^\dagger] \ket{\mathcal{S}^{\text{pyr.}}_{n,m}} = 0$.

\subsection{Other towers of states in the ``pyramid"}
\label{sec:othertowers}
In addition to allowing us to move between states $\ket{\mathcal{S}^\text{pyr.}_{n,0}} = \ket{\mathcal{S}^{IS}_{n}}$, the operator $R^\dagger$ in fact also allows us to move between states $\ket{\mathcal{S}^\text{pyr.}_{n,1}}$:
\begin{equation}
    \ket{\mathcal{S}^\text{pyr.}_{n,1}} = (R^\dagger)^{n-1} \ket{\mathcal{S}^\text{pyr.}_{1,1}}~.
    \label{eq:BSn}
\end{equation}
This gives an alternate proof for $\ket{\mathcal{S}^\text{pyr.}_{n,1}}$, which is immediate from the commutator in Eq.~(\ref{eq:Hlambdacomm}). Since we know that $\ket{\mathcal{S}^\text{pyr.}_{1,1}}$ is an eigenstate, we just have to show that the commutator annihilates all $\ket{\mathcal{S}^\text{pyr.}_{n,1}}$. To do so we simply write:
\begin{align}
   \label{eq:HlambdacommBS}
    &\left[H_\lambda, R^\dagger \right]\ket{\mathcal{S}^\text{pyr.}_{n,1}}/(2\lambda) \\
    &= \sum_{j=1}^L (-1)^j (\ketbra{1010}{1100}-\ketbra{0101}{0011})_{j-1,...,j+2}\ket{\mathcal{S}^\text{pyr.}_{n,1}}\nonumber\\
    &= \sum_{j=1}^L (-1)^j \!\ket{01010}\!\!\big(\!\bra{01100}\!+\!\bra{00110}\!\big)_{j-2,..,j+2}\!\ket*{\mathcal{S}^\text{pyr.}_{n,1}}\! =\! 0. \nonumber
\end{align}
In going to the last line we used that in $\ket{\mathcal{S}^\text{pyr.}_{n,1}}$, the sequence `11' is always surrounded by `0's. Finally, the `11' bound magnon has wavenumber $k=\pi$, hence the expression is 0. [More algebraically, $\ket{\mathcal{S}^\text{pyr.}_{1,1}}$ is in the common null space of all $\ket{...}\!\!\big(\!\bra{01100}\!+\!\bra{00110}\!\big)_{j-2,..,j+2}$, and this null space is preserved by the action of $R^\dagger$.]

The same expression Eq.~(\ref{eq:BSn}) does not work for $\ket{\mathcal{S}^\text{pyr.}_{n,m}}$ with $m>1$ because there are more types of domains than simply `1' and `$\mathbf{1}^{m+1}$'. This is also related to the fact that the ``ladder operators" $R^\dagger$ and $\mathcal{P}^\dagger$ do not commute. However, we note that we trivially have, for $m>1$:
\begin{align}
    \ket{\mathcal{S}^\text{pyr.}_{n,m}} &= \left((\mathcal{P}^\dagger)^m R^\dagger (\mathcal{P}'^\dagger)^m\right)^{n-1} \ket{\mathcal{S}^\text{pyr.}_{1,m}}\\
    &= \left((\mathcal{P}^\dagger)^{m-1} R^\dagger (\mathcal{P}'^\dagger)^{m-1}\right)^{n-1} \ket{\mathcal{S}^\text{pyr.}_{1,m}}~, \nonumber
\end{align}
where we used that $\mathcal{P}'^\dagger$ undoes the action of $\mathcal{P}^\dagger$ on the states in the pyramid and used $R^\dagger$ to move upwards either along the side of the pyramid or along one level below it.

\section{Perfect revivals from initial states}
\label{sec:perfectrevivals}
In this section we briefly review perfect revivals in such systems with exact scar towers. In Ref.~\cite{schecter_weak_2019}, Schecter and Iadecola identified an initial state $\ket{\psi_0}$ which, when quenched from, gives perfect revivals under their spin-1 XY model:
\begin{equation}
    \ket{\psi_0} = \otimes_j \left(\frac{\ket{1}_j - e^{i \mathbf{r}_j \cdot \pmb{\pi}} \ket{-1}_j}{\sqrt{2}} \right) ~.
\end{equation}
This is the state of minimal $J^x=-V/2$ value in the $J=V/2$ multiplet and as such is a superposition of the scar tower states $\ket{\mathcal{S}_n^{XY}}$. This state can be prepared as the ground state of the Hamiltonian $H = Q^\dagger + Q \propto J^x$. Since these states are equally spaced in energy with spacing $2h$, time evolving $\ket{\psi_0}$ gives perfect revivals with frequency $2h$.

For the AKLT tower of scars, while we do not know of such a ``rotation" in the space of scar states that produces a simple initial state, we can compress the scar tower into a state with finite MPS bond dimension, by taking:
\begin{equation}
    \ket{\psi_0^A} = \exp(z Q^\dagger) \ket{G} = \sum_n \frac{z^n}{n!} \ket{\mathcal{S}_{2n}} ~,
\end{equation}
with parameter $z$ that can take arbitrary value. By writing $\exp(z Q^\dagger)$ as a
bond-dimension 2 matrix product operator (MPO) (assuming even $L$ throughout):
\begin{align}
    &\exp(z Q^\dagger) = \prod_{j=1}^L \left[1 + (-1)^j z (S_j^+)^2 \right] 
    = b_l^T \left(\prod_{j=1}^L M_j \right) b_r ~,\\
    &M_j = \begin{pmatrix}
    -\mathbf{I}_j & -z(S^+_j)^2 \\
    z(S^+_j)^2 & \mathbf{I}_j
    \end{pmatrix},~~
    b_l = \begin{pmatrix}
    1 \\
    0
    \end{pmatrix},~~
    b_r = \begin{pmatrix}
    1 \\
    1
    \end{pmatrix}, \nonumber
\end{align}
we can write $\ket{\psi_0^A}$ as an MPO$\times$MPS~\cite{moudgalya_entanglement_2018}, itself an MPS. This has a bond dimension of $2\times4 = 8$ for the AKLT scar tower in PBC. Indeed, when we convert $\ket{G}$ in PBC from its trace expression in Eq.~(\ref{eq:MPS}) into a boundary vector form, $\ket{G}$ can be expressed as an MPS with bond-dimension 4 matrices $A^{[\sigma]}\oplus A^{[\sigma]}$ and boundary vectors $v_l = v_r =  \begin{pmatrix} 1 & 0 & 0 & 1
\end{pmatrix}^T$.
(The AKLT ground state in OBC would be already in such a boundary vector form with bond dimension 2, and the corresponding MPS compression of the AKLT tower in OBC would have bond dimension 4.) While we do not know of a ``simple" parent Hamiltonian with $\ket{\psi_0^A}$ as ground state, any finite bond-dimension MPS has a local parent Hamiltonian~\cite{fannes_finitely_1992,fannes_abundance_1992,perez-garcia_matrix_2007}, which could be possibly realized in cold atom systems, for example.

In the spin-1/2 system, Iadecola and Schecter~\cite{iadecola_quantum_2019} describe an initial state
\begin{equation}
    \ket{\xi} = \exp(\xi R^\dagger)\ket{\Omega}~.
\end{equation}
This has parent Hamiltonian in the Hilbert space of configurations with no adjacent 1s (the so-called ``Rydberg-blockaded" space): \begin{equation}
    H_\text{Parent} = \sum_j P^0_{j-1} [\xi^{-1} P^1_{j} + \xi P^0_j - (-1)^j \sigma^x_j] P^0_{j+1}~,
    \label{eq:sp12parent}
\end{equation}
which is related to the Lesanovksy model in Rydberg gases~\cite{Lesanovsky2011} by a similarity transformation. While it is not immediately clear how to obtain initial states for general combinations of ``pyramid" states, from the expression in Eq.~(\ref{eq:BSn}), we can also compress the states $\ket*{\mathcal{S}^\text{pyr.}_{n,1}}$ in an identical way with the initial state $\ket{\xi_1} = \exp(\xi R^\dagger)\ket*{\mathcal{S}^\text{pyr.}_{1,1}}$.

This initial state is in fact also a ground state of $H_\text{Parent}$ in Eq.~(\ref{eq:sp12parent}), but in the space where there is exactly one pair of adjacent `1's (the space with $N_{11} = 1$, where $N_{11} = \sum_j \ketbra{11}_{j,j+1}$). $H_\text{Parent}$ not only preserves the number of Rydberg violations, it does nothing on the `11' block. Therefore $\ket{\xi_1}$ is not the unique ground state of $H_\text{Parent}$. Rather, the ground state manifold is spanned by the states $\exp(\xi R^\dagger) \ket{0 \cdots 0 1_{j}1 0 \cdots 0}$, $j=1,...,L$. This is most easily seen by the fact that the `11' block effectively turns the problem into one in the space with $N_{11} = 0$, with $L-4$ sites and open boundary conditions. 

We can make $\ket{\xi_1}$ a unique ground state by adding the `11' block hopping-type term $\sum_j (\ket{0_j1100}+\ket{0_j0110})(\bra{0_j1100}+\bra{0_j0110})$. $\ket{\xi_1}$ is the only state in the previous ground state manifold that is annihilated by this term and becomes the unique ground state. It is also easy to verify that we have the desired commutation relation $\left[\sum_j (\ket{0_j1100}+\ket{0_j0110})(\bra{0_j1100}+\bra{0_j0110}),R^\dagger\right]\!\ket*{\mathcal{S}^\text{pyr.}_{n,1}}\allowbreak= 0$. However, because accessing the $N_{11} = 1$ space is not experimentally possible in Rydberg-blockaded systems, this parent Hamiltonian is likely only of theoretical interest.

Finding suitable initial states and parent Hamiltonians for compressions along other cuts on the pyramid is another possible direction of future work.

\section{Conclusion}
Using a common framework of showing that the commutator $\left[H, Q^\dagger \right]$ annihilates states in a special subspace, we have provided simple proofs for the exact towers of states in the 1D AKLT models, spin-1 XY model and the domain-wall-conserving spin-1/2 model. This not only clarifies the structure of these scar states but also allows us to immediately provide a family of Hamiltonians that share the AKLT scar states. We also introduce new exact towers in the spin-1/2 model, which are organized in a pyramid-like structure. This shows that the spin-1/2 model hosts many more exact states in PBC than it was thought to, and enhances our understanding of the known scar tower.

Answering the question of whether the discussed scar tower models can be cast in Shiraishi-Mori form is an interesting problem for future work. The commutator framework also hints at a generalization of the Shiraishi-Mori structure to engineer Hamiltonians which host scar towers and more generally improve our understanding of quantum many-body scars. Realizing the ingredients of our simple theorem in Eq.~(\ref{eq:asimpletheorem}) presents a heuristic for producing models with scar towers. We note that a very recent paper~\cite{shibata_onsagers_2019} does a subcase of this with their operator $Q^+$ being essentially a symmetry of a base Hamiltonian, atop which they add perturbations that annihilate a tower of scar states.
In Sec.~\ref{sec:parentham}, we discuss how the scars in the spin-1 AKLT and XY models are related to an $SU(2)$ pseudospin symmetry generated by their ladder operator $Q^\dagger$. The $SU(2)$-invariant ``parent Hamiltonian" $H_0$ is embedded in both models and the $SU(2)$ invariance is destroyed by perturbing terms that annihilate the respective scar towers.

While in this paper we focused on exact eigenstates in the subspace $W$ identified in Eq.~(\ref{eq:asimpletheorem}), it is also natural to ask about the thermalization dynamics of other states in $W$. The AKLT model exhibits such a space $W$ that is larger than its scar tower, and could be an interesting subject for future work.

Finally, we highlight our finding of the pyramid structure of scar states in the domain-wall-preserving spin-1/2 chain. We have not encountered such a structure before, and it would be interesting to study properties of all states in the pyramid, develop better understanding of aspects of the model responsible for it, and also look for other models realizing such scars.

\begin{acknowledgments}
We thank Alvaro Alhambra, Anushya Chandran, Timothy Hsieh, Rahul Nandkishore, Tibor Rakovszky, and Christopher Turner for valuable discussions. We also thank Andrei Bernevig, Juan P. Garrahan, Thomas Iadecola, Hosho Katsura, Sanjay Moudgalya, Zlatko Papi\'c, and Nicolas Regnault for comments and discussions on the manuscript.
D.~M.\ acknowledges funding from the James C.~Whitney SURF Fellowship, Caltech Student-Faculty Programs. This work was also supported by National Science Foundation through Grant DMR-1619696.
C.-J.~L.\ acknowledges support from Perimeter Institute for Theoretical Physics.
Research at Perimeter Institute is supported in part by the Government of Canada through the Department of Innovation, Science and Economic Development Canada and by the Province of Ontario through the Ministry of Economic Development, Job Creation and Trade.
\end{acknowledgments}

\appendix

\section{Two-site spin states in the spin-1 AKLT model}
\label{sec:twositestates}
Here we list the states of total spin $J$ and magnetization $M$ defined on spin-1's on two sites. We use these states extensively in our discussion of the scar states in the spin-1 AKLT model in Section~\ref{sec:spin1AKLT}:
\begin{gather}
    \ket{T_{2,-2}} = \ket{-1,-1}~,~~ \ket{T_{2,-1}} = \frac{1}{\sqrt{2}} \left(\ket{0,-1} + \ket{-1,0} \right)~, \nonumber\\
    \ket{T_{2,0}} = \frac{1}{\sqrt{6}}\left(\ket{1,-1} + 2\ket{0,0} + \ket{-1,1} \right)~, \nonumber\\
    \ket{T_{2,1}} = \frac{1}{\sqrt{2}}\left(\ket{1,0} + \ket{0,1} \right)~,~~ \ket{T_{2,2}} = \ket{1,1}, \nonumber\\
    \ket{T_{1,-1}} = \frac{1}{\sqrt{2}}\left(\ket{0,-1} - \ket{-1,0}\right)~, \label{eq:twositestates}\\
    \ket{T_{1,0}} = \frac{1}{\sqrt{2}}\left(\ket{1,-1} - \ket{-1,1} \right)~, \nonumber\\
    \ket{T_{1,1}} = \frac{1}{\sqrt{2}}\left(\ket{1,0} - \ket{0,1} \right)~, \nonumber\\
    \ket{T_{0,0}} = \frac{1}{\sqrt{3}}\left(\ket{1,-1} - \ket{0,0} + \ket{-1,1} \right) ~.\nonumber
\end{gather}

\section{Proof of additional scar tower in the spin-1 XY model}
\label{sec:XYProof}
Schecter and Iadecola numerically found and conjectured an additional scar tower $\ket{\mathcal{S}^{XY,2}_n}$ that is only present in 1D with $J_3=D=0$.
This tower was subsequently proven in by Chattopadhyay \textit{et. al.}~\cite{chattopadhyay_quantum_2019} by compressing the states into a bond dimension two MPS state and using MPS techniques to show that the variance of $H_{XY}$ vanishes in this state. Here we present an alternative proof through direct evaluation of $H_{XY} \ket{\mathcal{S}^{XY,2}_n}$.
We first note that $H_z \ket{\mathcal{S}^{XY,2}_n} = h(2n-L) \ket{\mathcal{S}^{XY,2}_n}$ and the perturbation $V$ in Ref.~\cite{chattopadhyay_quantum_2019} annihilates these scar states.  We then have to show that Eq.~(\ref{eq:ISXYscar2}) is an eigenstate of $H_{XY}$, specifically that $H_{XY} \ket{\mathcal{S}^{XY,2}_n} = 0$, where:
\begin{gather}
    \ket{\mathcal{S}^{XY,2}_n} = \mathcal{O}_n \ket{\Omega},~~~ \ket{\Omega} = \ket{-1,-1,\cdots,-1}, \\
    \mathcal{O}_n = \sum_{i_1 \neq \cdots \neq i_n} \!\!\!\!(-1)^{\sum_j i_j} (S_{i_1}^+ S_{i_1+1}^+)\cdots (S_{i_n}^+ S_{i_n+1}^+)~. \nonumber
    \label{eq:hxyO}
\end{gather}
To do so we trivially require that $i_1 < \cdots < i_n$ and write:
\begin{gather}
    H_{XY} \mathcal{O}_n \ket{\Omega} = \sum_{i_1 < \cdots < i_n} \left(\sum_{l \in \mathcal{L}} h_{l} \right) (-1)^{\sum_j i_j} s_{i_1} \cdots s_{i_n} \ket{\Omega}, \nonumber \\
    \mathcal{L} = \bigcup_{j=1}^n \left\{i_j-1,i_j, i_j+1 \right\} ~,
\end{gather}
where we have abbreviated $s_{i_j} = S_{i_j}^+ S_{i_j+1}^+$ and $h_l = \left(S_l^+ S_{l+1}^- + S_l^- S_{l+1}^+ \right)/2$. The sum over $l$ in $\mathcal{L}$ keeps only the nontrivial Hamiltonian terms. 
We can then split the sum as follows:
\begin{align}
    H_{XY} \mathcal{O}_n \ket{\Omega}& \label{eq:Oexpand} \\
    = \sum_{j=1}^n \sum_{\{i_k, k \neq j\}} &\left(\sum_{i_j = i_{j-1} + 1}^{i_{j+1} - 1} \!\!\! (-1)^{i_j} g_{i_j}\left(i_{j-1},i_j,i_{j+1}\right) s_{i_j} \right) \nonumber\\
    &\times\prod_{k \neq j} (-1)^{i_k}  s_{i_k} \ket{\Omega}, \nonumber 
\end{align}
where the $g_{i_j}\left(i_{j-1},i_j,i_{j+1}\right)$ denotes that we assign a group of Hamiltonian terms in $\mathcal{L}$ to each of the $i_j$ in such a way that it avoids double counting. This grouping can depend also on the relative positions of $i_{j-1}$ and $i_{j+1}$. Specifically, it always includes $h_{i_j}$ but also includes full, parts, or none of $h_{i_j-1}$ and $h_{i_j+1}$, depending on the distance of $i_j$ to $i_{j\pm1}$. We outline the groupings below and state it explicitly in Eq.~(\ref{eq:htildeXYterms}). 

The motivation for such a rewriting is to isolate each $i_j$. The term in the parentheses is then a sum over all possible $i_j$, keeping all other $i_k$ fixed. We will prove that $H_{XY} \mathcal{O}_n \ket{\Omega} = 0$ by showing that each bracketed term in Eq.~(\ref{eq:Oexpand}) gives 0, via pairwise cancellations of contributions to the sum inside the bracket.

We start by listing the configurations in $\mathcal{O}_n\ket{\Omega}$. Fixing $i_k, k\neq j$, $i_j$ ranges between $i_{j-1}+1$ and $i_{j+1}-1$. In the below table we list the possible configurations, indicating the positions of $i_{j-1},i_{j},i_{j+1}$ by underlining them.
\begin{equation}
\begin{tabular}{c|| c r | l c r| l}
 $i_j$ & ~~sign~~ & \multicolumn{5}{c}{$(-1)^{i_j}s_{i_{j-1}}s_{i_j} s_{i_{j+1}} \ket{\Omega}$}  \\
$i_{j-1}+1$&($\pm$)~&$\overbracket{\,\underline{0\underline{+}}}^{i_{j-1}}\!$&$\!\underline{\vphantom{+}\,0}\!-\!-$ & $\cdots$ & $-\!-\!-$ & \!$\overbracket{\underline{00}}^{i_{j+1}}$ \\
$i_{j-1}+2$&($\mp$)&$\underline{00}$& $\underline{0\,0\,}-$ & $\cdots$ & $-\!-\!-$ & $\underline{00}$\\
$i_{j-1}+3$&($\pm$)&$\underline{00}$&$\!-\underline{\,0\,0}$ & $\cdots$ & $-\!-\!-$ & $\underline{00}$\\
\vdots& & & & $\vdots$ & &\\
$i_{j+1}-3$& $(\pm')$&$\underline{00}$&$-\!-\!-$& $\cdots$ & $\underline{\,0\,0\,}-$ &$\underline{00}$\\
$i_{j+1}-2$& $(\mp')$&$\underline{00}$&$-\!-\!-$& $\cdots$ & $-\underline{\,0\,0\,}$ &$\underline{00}$\\
$i_{j+1}-1$& $(\pm')$&$\underline{00}$&$-\!-\!-$& $\cdots$ & $-\!-\!\underline{\vphantom{+}0\,}$\! &$\!\underline{\underline{+}0}$
\end{tabular}
\label{eq:HXYconfigstable}
\end{equation}
In this section we use `+'/'-' to denote spin-1 states `+1'/`-1' for concision. We also note that though we write the left and right boundaries as always `0', they can be `+' without affecting the argument. The `sign' column is meant to indicate that the specific sign factor $(-1)^{i_j}$ changes sign for each increase of $i_j$ by one; this will be particularly convenient for noticing cancellations when examining the action of the terms in the Hamiltonian on these configurations.

For each configuration position $i_j$, $h_k$ is only nonzero when $k = i_j-1,i_j,$ or $i_j +1$. We only keep specific terms that interact with $s_{i_j}$ to avoid double counting. We first consider the action of the terms $h_{i_j}$, omitting the constant $J$:
\begin{equation}
\begin{tabular}{c|| c r | l c r | l}
 $i_j$&~~sign~~& \multicolumn{5}{c}{$(-1)^{i_j}h_{i_j}s_{i_{j-1}}s_{i_j} s_{i_{j+1}}\ket{\Omega}$}  \\
$i_{j-1}+1$&($\pm$)~&$\underline{0\underline{\vphantom{+}0}}$\!&$\!\underline{+}\!-\!-$ & $\cdots$ & $---$ & $\underline{00}$ \\
$i_{j-1}+2$&($\mp$)&$\underline{00}$\!& $\!\underline{+-}-$ & $\cdots$ & $-\!-\!-$ & $\underline{00}$\\
$i_{j-1}+2$&($\mp$)&$\underline{00}$&$\!\underline{-+}-$ & $\cdots$ & $-\!-\!-$ & $\underline{00}$\\
$i_{j-1}+3$&($\pm$)&$\underline{00}$&$\!-\underline{+-}$ & $\cdots$ & $-\!-\!-$ & $\underline{00}$\\
$i_{j-1}+3$&($\pm$)&$\underline{00}$&$\!-\underline{-+}$ & $\cdots$ & $-\!-\!-$ & $\underline{00}$\\
\vdots& & & & $\vdots$ & &\\
$i_{j+1}-3$& $(\pm')$&$\underline{00}$&$-\!-\!-$& $\cdots$ & $\underline{+-}-$ &$\underline{00}$\\
$i_{j+1}-3$& $(\pm')$&$\underline{00}$&$-\!-\!-$& $\cdots$ & $\underline{-+}-$ &$\underline{00}$\\
$i_{j+1}-2$& $(\mp')$&$\underline{00}$&$-\!-\!-$& $\cdots$ & $-\underline{+-}$ &$\underline{00}$\\
$i_{j+1}-2$& $(\mp')$&$\underline{00}$&$-\!-\!-$& $\cdots$ & $-\underline{-+}$ &$\underline{00}$\\
$i_{j+1}-1$& $(\pm')$&$\underline{00}$&$-\!-\!-$& $\cdots$ & $-\!-\!\underline{+}$\! &$\!\underline{\underline{\vphantom{+}0}0}$
\end{tabular}
\label{eq:apphjSj}
\end{equation}
As in Eq.~(\ref{eq:HXYconfigstable}) we denote the positions of $i_{j-1}$, $i_j$, and $i_{j+1}$ by underlining the relevant two sites. Two rows with the same $i_j$ appear from nonzero action of both terms in each $h_{i_j}$ and are listed in convenient order. We see that the successive rows above cancel. 

We next consider the terms with $h_{i_j\pm1}$ that we include in $g_{i_j}\left(i_j-1, i_j, i_j+1 \right)$:
\begin{equation}
\begin{tabular}{c || c r | l c r | l  l}
 \multicolumn{8}{c}{$(-1)^{i_j}\left(h_{i_j-1} + h_{i_j+1}\right)s_{i_{j-1}}s_{i_j} s_{i_{j+1}}\ket{\Omega}$}  \\
 $i_j$&~~sign~~&&&&&&operator\\
$i_{j-1}+1$&~~($\pm$)~~~&$\underline{0\underline{+}}\!$&$\!\underline{\,\vphantom{+}-}0\,-$ & $\cdots$ & $-\!-\!-$ & $\underline{00}$ & $(h_{i_{j}+1})$*\\
$i_{j-1}+2$&($\mp$)&$\underline{0+}$\!& $\underline{-0}-$ & $\cdots$ & $-\!-\!-$ & $\underline{00}$ & $(h^{+-}_{i_{j}-1})$**\\
$i_{j-1}+2$&($\mp$)&$\underline{00}$&$\underline{0-}0$ & $\cdots$ & $-\!-\!-$ & $\underline{00}$ & $(h_{i_{j}+1})$\\
$i_{j-1}+3$&($\pm$)&$\underline{00}$&$ 0\underline{-0}$ & $\cdots$ & $-\!-\!-$ & $\underline{00}$ &$(h_{i_{j}-1})$\\
$i_{j-1}+3$&($\pm$)&$\underline{00}$&$ -\underline{0-}$ & $\cdots$ & $-\!-\!-$ & $\underline{00}$ &$(h_{i_{j}+1})$\\
\multicolumn{1}{c||}{\vdots} & & & & $\vdots$ & &\\
$i_{j+1}-3$& $(\pm')$&$\underline{00}$&$-\!-\!-$& $\cdots$ & $\underline{-0}-$ &$\underline{00}$&$(h_{i_{j}-1})$\\
$i_{j+1}-3$& $(\pm')$&$\underline{00}$&$-\!-\!-$& $\cdots$ & $\underline{0-}0$ &$\underline{00}$&$(h_{i_{j}+1})$\\
$i_{j+1}-2$& $(\mp')$&$\underline{00}$&$-\!-\!-$& $\cdots$ & $0\underline{-0}$ &$\underline{00}$&$(h_{i_{j}-1})$\\
$i_{j+1}-2$& $(\mp')$&$\underline{00}$&$-\!-\!-$& $\cdots$ & $-\underline{0-}$ &$\!\underline{+0}$&$(h^{-+}_{i_{j}+1})$**\\
$i_{j+1}-1$&$(\pm')$&$\underline{00}$&$-\!-\!-$& $\cdots$ & $-0\underline{\vphantom{+}-\,}$\! &$\!\underline{\underline{+}0}$&$(h_{i_{j}-1})$*
\end{tabular}
\label{eq:apphjmpSj}
\end{equation}
\newline
Again, we denote the positions of $i_{j-1}$, $i_j$, and $i_{j+1}$ by underlining the relevant two sites [so the result of $(-1)^{i_j} s_{i_{j-1}} s_{i_j} s_{i_{j+1}} \ket{\Omega}$ can be read from Eq.~(\ref{eq:HXYconfigstable})], while we indicate which of the $h$ terms acts in the last column.  Note that near the boundaries we only include parts of the $h$ terms to avoid double counting, which is marked and explained as follows: The * indicates that when $i_{j} = i_{j\pm1} \mp 1$, the $h_{i_{j}\pm1}s_{i_j}$ term is left out to avoid double counting because it appears in the sum Eq.~(\ref{eq:apphjSj}) (for $i_{j\pm1}$ instead of $i_j$) as $h_{i_{j\pm1}}s_{i_{j\pm1}}$. The ** indicates that when $i_{j} = i_{j\pm1} \mp 2$, in the term $h_{i_{j}\pm1}s_{i_j}$, to avoid double counting with Eq.~(\ref{eq:apphjmpSj}) (for $i_{j\pm1}$ instead of $i_{j}$), we only keep the $ h^{+-}_{i_{j-1}+1} = S^{+}_{i_{j-1}+1}S^{-}_{i_{j-1}+2}/2$ and $h^{-+}_{i_{j+1}-1} = S^{-}_{i_{j+1}-1}S^{+}_{i_{j+1}}/2$ terms (the ones where the minus is ``facing inwards"). 
We also point out that the above table is correct for $i_{j+1} - i_{j-1} \geq 4$. For $i_{j+1} - i_{j-1} < 4$ boundary effects at both ends have to be considered for each term, which follows the general formula for $g_{i_j}\left(i_{j-1},i_j,i_{j+1}\right)$ given in Eq.~(\ref{eq:htildeXYterms}); one can construct similar table for each such case as well. Having taken care not to double count any terms, we see that the successive rows in Eq.~(\ref{eq:apphjmpSj}) cancel, and this proves the desired $H_{XY}\ket{\mathcal{S}^{XY,2}_n} = 0$.

To summarize, the grouping $g_{i_j}\left(i_j-1, i_j, i_j+1 \right)$ is as follows:
\begin{align}
&g_{i_j}\left(i_j-1, i_j, i_j+1 \right) = \label{eq:htildeXYterms} \\
&h_{i_j} +
    \left\{\begin{aligned}
    &h_{i_j-1},~i_j > i_{j-1} + 2 \\
    &h^{+-}_{i_j-1},~i_j = i_{j-1} + 2 \\
    &0,~i_j = i_{j-1} + 1
    \end{aligned}\right\}
    +
    \left\{\begin{aligned}
    &h_{i_j+1},~i_j < i_{j+1} - 2 \\
    &h^{-+}_{i_j+1},~i_j = i_{j+1} - 2 \\
    &0,~i_j = i_{j+1} - 1
    \end{aligned}\right\}~. \nonumber
\end{align}
This accounts for all instances of double counting and gives the desired partitioning of $H_{XY}$ terms into $\tilde{h}_{i_j}$ in Eq.~(\ref{eq:Oexpand}).

\section{Connection of SU(2)-invariant ``parent Hamiltonian" $H_0$ to an integrable spin-1 chain}
\label{sec:H0appendix}
In Sec.~\ref{sec:parentham}, we introduced a ``common parent" Hamiltonian $H_0$ that contained the scar states $\ket{\mathcal{S}_{2n}}$ and $\ket{\mathcal{S}_{n}^{XY}}$ of the spin-1 AKLT and XY models. Here we discuss its connection to an integrable spin-1 model.

For reference we restate $H_0$, defined on any bipartite graph:
\begin{equation}
    H_0 = \sum_{\expval{ij}} \big(\ketbra{1,0}{0,1} - \ketbra{-1,0}{0,-1} + \hc \big)_{i,j} ~.
\end{equation}

Denoting the bipartite subsets of the graph as $A$ and $B$, $H_0$ can be unitarily transformed to the following model:
\begin{align}
    \tilde{H}_0 &= U H_0 U ~,~~ U = \bigotimes_{j\in A} \begin{pmatrix} 1 & 0 & 0 \\ 0 & 1 & 0 \\ 0 & 0 & -1 \end{pmatrix}_j
    ~, \\
    \tilde{H}_0 &= \sum_{\expval{ij}} \big(\ketbra{1,0}{0,1} + \ketbra{-1,0}{0,-1} + \hc \big)_{i,j} ~. \nonumber
\end{align}
The unitary $U$ assigns, on one subset $A$, a coefficient of $-1$ on $\ket{-1}$ and $1$ on $\ket{1}$ and $\ket{0}$ (and acts as an identity on the other subset $B$). $\tilde{H}_0$ can be viewed as a hopping problem of spin-1/2 hard core bosons with no double occupancy, with identification $\ket{1} \equiv b_\uparrow^\dagger \ket{\text{vac}_b}$, $\ket{0} \equiv \ket{\text{vac}_b}$, $\ket{-1} \equiv b_\downarrow^\dagger \ket{\text{vac}_b}$, while the correspondingly transformed $(\tilde{J}^{\pm}, \tilde{J}_z)$ generate the standard global SU(2) symmetry of $b_\uparrow, b_\downarrow$, which is manifest in $\tilde{H}_0$ in this language.

In 1D, $\tilde{H}_0$ is in fact integrable~\cite{klumper_exact_1995, mutter_solvable_1995}, and a more general family of models is solved in Ref.~\cite{klumper_exact_1995} through a mapping to the six-vertex model. (For reference, $\tilde{H}_0$ is equivalent to model $H_4$ in Table 3 in Ref.~\cite{mutter_solvable_1995} and corresponds to a special case of models III and IV in Ref.~\cite{klumper_exact_1995}.)

We can actually solve the model $\tilde{H}_0$ directly by more elementary means.  We first note that $N_1$, $N_0$, and $N_{-1}$ are separately conserved and that all states with $N_0 = 0$ are annihilated by $\tilde{H}_0$.  The model in the sectors with $N_{-1} = 0$ essentially corresponds to a hard-core boson hopping model in terms of $b_\uparrow$ and is solved by the standard Jordan-Wigner transformation to free fermions, and similarly in the sectors with $N_1 = 0$.  We next note that $\tilde{H}_0$ preserves the pattern of `$\pm 1$'s, i.e., the ordering of `$\pm 1$'s on the chain irrespective of the intervening `$0$'s, where for simplicity we specialize to open boundary conditions.  Then all solutions in a sector with given $(N_1, N_0, N_{-1}) = (a, b, c)$ can be obtained by acting with $(\tilde{J}^{+})^a$ on the known solutions in the sector $(0, b, c+a)$ and splitting the result into subsectors defined by the `$\pm 1$' patterns. Every sector of a fixed `$\pm1$' pattern thus has an identical spectrum as the sector with no `$+1$'s and the same number of `0's. In PBC, each sector is defined instead by equivalence classes of `$\pm1$' patterns, equivalent under translation of the pattern, and more care has to be taken to solve $\tilde{H}_0$, as discussed in Ref.~\cite{caspers_exact_1989}.

It is worth emphasizing, however, that the embedding of the AKLT and spin-1 XY model towers in $H_0$ in Sec.~\ref{sec:parentham} did not require any knowledge of the solvability of $H_0$.  In particular, the embedding analysis goes through identically also for the nonintegrable modification of $H_0$ discussed at the end of Sec.~\ref{sec:parentham}.

\section{Commutator proof to ``pyramid" of states in the spin-1/2 model}
\label{sec:prooftoorthogonal}
A proof of $\ket{\mathcal{S}^\text{pyr.}_{n,m}} = (\mathcal{P}^\dagger)^m \ket{\mathcal{S}_n}$ was given in Section~\ref{sec:directprooforthogonal}. Here we provide an alternate proof by showing that $[H_\lambda, \mathcal{P}^\dagger]\ket{\mathcal{S}^\text{pyr.}_{n,m}} = 0$, where we remind for convenience
\begin{align*}
    \mathcal{P}^\dagger &=\sum_{j=1}^L \sum_{l=1}^{L-2} P^{\mathbf{1},l}_{j-1} \sigma_j^+ P^0_{j+1} \\
    &= \sum_{j=1}^L P^{1}_{j-1} \sigma_j^+ P^0_{j+1} + P^{1}_{j-2} P^{1}_{j-1} \sigma_j^+ P^0_{j+1} + ... ~,\nonumber
\end{align*}
and $H_\lambda = 2\lambda \sum_{j=1}^L \left(P^1_{j-1} \sigma^x_j P^0_{j+1} + P^0_{j-1} \sigma^x_j P^1_{j+1} \right)$. We calculate the various terms as follows:
\begin{gather}
    \sum_j \left[H_\lambda, P^{1}_{j-1} \sigma_j^+ P^0_{j+1} \right] = 2\lambda \sum_j \Big(-P^1_{j-1} \sigma^z_j P^0_{j+1} \nonumber \\
     + P^0_{j-2} \sigma^-_{j-1} \sigma^+_j P^0_{j+1} - P^1_{j-1} \sigma^+_j \sigma^-_{j+1} P^1_{j+2} \Big) \nonumber\\
    = 2\lambda \sum_j \Big(\left(\ket{0100} + \ket{0010} \right)\!\bra{0100}~~~~~~ \nonumber\\
    - \ket{1101}\!\left(\bra{1101} + \bra{1011}\right) \Big)_{j,...,j+3} ~,
\label{eq:range1comm}
\end{gather}
where in going to the last line we have expanded:
\begin{gather}
    -\sum_j P^1_{j-1} \sigma^z_j P^0_{j+1} = \sum_j \Big((P^1 + P^0)_{j-2} P^1_{j-1} P^0_j P^0_{j+1} \nonumber \\
     - P^1_{j-1} P^1_j P^0_{j+1} (P^1 + P^0)_{j+2} \Big) \nonumber \\
    = \sum_j \left(\ketbra{0100} - \ketbra{1101} \right)_{j,...,j+3} ~.
    \label{eq:sigzexpand}
\end{gather}
Evaluating the commutator with the other terms is not much more difficult because $H_\lambda$ is only nontrivial on a few sites near the left and right boundaries of each term in $\mathcal{P}^\dagger$. For all $l>1$:
\begin{gather}
    \sum_j \left[H_\lambda, P^{\mathbf{1},l}_{j-1} \sigma_j^+ P^0_{j+1} \right] = \nonumber \\
    2\lambda \sum_j \Big(-P^{\mathbf{1},l}_{j-1} \sigma^z_j P^0_{j+1} + P^0_{j-l-1} (\sigma^x \sigma^z)_{j-l} P^{\mathbf{1},l-1}_{j-1} \sigma^+_j P^0_{j+1} \nonumber \\
    - P^0_{j-l} P^{\mathbf{1},l-1}_{j-1} \sigma^+_j \sigma^+_{j+1} P^0_{j+2} - P^{\mathbf{1},l}_{j-1} \sigma_j^+ \sigma^-_{j+1} P^1_{j+2} \Big) \nonumber \\
    = 2\lambda \sum_j \Big( (\ket*{00\mathbf{1}^{l}0} + \ket*{0\mathbf{1}^{l}00})\bra*{0\mathbf{1}^{l}00} \nonumber \\
    - \ket*{0\mathbf{1}^{l+1}0}(\bra*{0\mathbf{1}^{l-1}000} + \bra*{00\mathbf{1}^{l-1}00}) \nonumber \\
    ~~~~~~~ - \ket*{\mathbf{1}^{l+1}01} (\bra*{\mathbf{1}^{l+1}01} + \bra*{\mathbf{1}^{l}011}) \Big)_{j,...,j+l+2} ~,
\end{gather}
where here and in the rest of the section we use the shorthand: $\ket{\mathbf{1}^l} = \ket*{\underbracket{1\cdots1}_l}$. Here we have also used a similar expansion as in Eq.~(\ref{eq:sigzexpand}) for the $\sigma^z_j$ term. Some care must be taken when $l=L-2$, where the boundary terms at $j$ and $j+l+2$ intersect. However, the commutator in this case can only act nontrivially on $\ket{\mathcal{S}^\text{pyr.}_{1,m}}$ with $m=L-3, L-2, L-1$, and it is easy to check that they are annihilated.

The term
\begin{equation}
-\ket*{0\mathbf{1}^{l+1}0}\!\big(\!\bra*{0\mathbf{1}^{l-1}000} + \!\bra*{00\mathbf{1}^{l-1}00}\big)_{j,...,j+l+2} ~,
\end{equation}
annihilates $\ket{\mathcal{S}^\text{pyr.}_{n,m}}$ because by construction each domain of `1's has wavenumber $k=\pi$.
Likewise, the term
\begin{equation}
-\ket*{\mathbf{1}^{l+1}01} (\bra*{\mathbf{1}^{l+1}01} + \!\bra*{\mathbf{1}^{l}011})_{j,...,j+l+2} ~,
\end{equation}
annihilates $\ket{\mathcal{S}^\text{pyr.}_{n,m}}$ because relevant contributions
in $\ket{\mathcal{S}^\text{pyr.}_{n,m}}$ can be grouped into pairs of configurations as follows: 
\begin{align*}
    (-1)^{m_1} \Big(-&\ket*{\cdots0\mathbf{1}^{m_1}0_{j+l+1}\mathbf{1}^{m_2}0\cdots} \\  
    +& \ket*{\cdots0\mathbf{1}^{m_1-1}0_{j+l}\mathbf{1}^{m_2+1}0\cdots}\Big) ~,
\end{align*}
which have opposite signs because of the different starting positions of the domains of `1's. This is also true for the $\ket{1101}(\bra{1101}+\bra{1011})$ term in Eq.~(\ref{eq:range1comm}).  Therefore, it remains only to consider the action of
\begin{equation}
    \mathcal{O} = \sum_{j=1}^L\sum_{l=1}^{L-3}  \!\!\left(\ket*{00\mathbf{1}^{l}0}\!+\!\ket*{0\mathbf{1}^{l}00}\right)\!\!\bra*{0\mathbf{1}^{l}00}_{j,..,j+l+2}~,
\end{equation}
on the pyramid states.

This operator does the following: whenever there are two `0's to the right of a domain of `1's, it shifts this domain one site to the right or keeps it in place. To show that this annihilates every $\ket{\mathcal{S}^\text{pyr.}_{n,m}}$, we use the expression Eq.~(\ref{eq:Snmexpansion}), repeated here:
\begin{equation*}
\ket{\mathcal{S}^\text{pyr.}_{n,m}} =
    \sum_{i_1 < i_2 < ... < i_n} \sum_{(l_j)} (-1)^{\sum_j i_j} \ket{i_1,...,i_n}_{(l_1,...,l_n)} ~.
\end{equation*}
For a given configuration with domain lengths $(l_j)$ and positions $\{i_j\}$,
\begin{gather}
    \mathcal{O} \ket{i_1,...,i_n}_{(l_1,...,l_n)} = n'  \ket{i_1,...,i_n}_{(l_1,...,l_n)} \nonumber \\
    + \sideset{}{'}\sum_j \ket{i_1,...i_j+1,...,i_n}_{(l_1,...,l_n)} ~,
    \label{eq:Pdaggcomm on sts.}
\end{gather}
where $n'$ denotes the number of domains of `1's that can be shifted one unit to the right, and the primed sum is over the $n'$ domain positions $i_j$. Summing over all $\{i_j\}$ with factor $(-1)^{\sum_j i_j}$ gives 0. This is because a configuration with $n'$ domains that can be shifted one unit to the right also has $n'$ domains that can be shifted one unit to the left, and so has exactly $n'$ preimages in the primed sum, which cancels out the $n'  \ket{i_1,...,i_n}_{(l_1,...,l_n)}$ term. This proves the desired $[H_\lambda, \mathcal{P}^\dagger] \ket{\mathcal{S}^\text{pyr.}_{n,m}} = 0$ and that $\ket{\mathcal{S}^\text{pyr.}_{n,m}}$ are eigenstates.

\section{Solution to states in the spin-1/2 model with $N_{DW}=2$}
\label{sec:NDW2}
In Sec.~\ref{sec:characterizingSnm} we provided exact expressions Eq.~(\ref{eq:Bm}) for exact states $\ket{\mathcal{S}^\text{pyr.}_{1,m}}$ with $N_{DW}=2$ and $k=\pi$. Here we discuss other states with $N_{DW}=2$. For each wavenumber $k = 2\pi n/L$, we form the basis states:
\begin{equation}
    \ket{\mathcal{B}_{m,k}} = e^{i k \frac{m}{2}} \sum_{j=1}^L e^{i k j} \ket*{0\cdots \underbracket{1_j\cdots1}_{m}\cdots0}.
\end{equation}
Then the action of $H_\lambda$ is, for $2 \leq m \leq L-2$:
\begin{equation}
    H_\lambda \ket{\mathcal{B}_{m,k}} = 4 \lambda \cos\left(\frac{k}{2} \right) (\ket{\mathcal{B}_{m-1,k}} + \ket{\mathcal{B}_{m+1,k}}) ~.
\end{equation}
 For $m = 1$ and $m = L-1$ hopping only occurs to $\ket{\mathcal{B}_{2,k}}$ and $\ket{\mathcal{B}_{L-2,k}}$ respectively. In this basis the Hamiltonian Eq.~(\ref{eq:HIS1/2}) is:
\begin{align}
    &H_{1/2}^{IS} = \left[J(L-4) - \Delta L\right] \mathbf{I}_{L-1} \\ 
    &+ 2\begin{pmatrix}
    \Delta & 2\lambda\cos(k/2) \\
    2\lambda\cos(k/2) & 2\Delta & \ddots \\
     & \ddots & (L-2)\Delta & 2\lambda\cos(k/2) \\
     &        & 2\lambda\cos(k/2) & (L-1)\Delta \\ 
    \end{pmatrix},\nonumber
\end{align}
where $\mathbf{I}_N$ is an $N \times N$ identity matrix. This Hamiltonian corresponds to a single-particle hopping problem on an OBC chain with linear potential and has been solved in Ref.~\cite{stey_wannier-stark_1973}. 
The eigenstates can be expressed in terms of Lommel polynomials and have energies
\begin{equation}
    E_j = 2\Delta (1 - \mu^{(L-1)}_j) + J(L-4) - \Delta L ~,
\end{equation}
for $j=1,...,L-1$, where $\mu^{(N)}_j$ is the $j$th zero of the Lommel polynomial $R_{N,\mu} (4\lambda\cos(k/2)/\Delta)$, considered as a function of $\mu$. For $k=\pi$, the hopping amplitude vanishes and the eigenstates are simply $\ket{\mathcal{B}_{m,\pi}}$ with energy $\Delta (2m-L) + J(L-4)$; they correspond to the states in the base of the pyramid of exact eigenstates in the main text, $\ket{\mathcal{S}^\text{pyr.}_{1,m-1}}$, see~Eq.~(\ref{eq:Bm}).

\end{document}